# Ramp: Fast Frequent Itemset Mining with Efficient Bit-vector Projection Technique

Shariq Bashir

National University of Computer and Emerging Sciences, Islamabad, Pakistan

shariq.bashir@nu.edu.pk

A. Rauf Baig

National University of Computer and Emerging Sciences, Islamabad, Pakistan

rauf.baig@nu.edu.pk

## Abstract

Mining frequent itemset using bit-vector representation approach is very efficient for dense type datasets, but highly inefficient for sparse datasets due to lack of any efficient bit-vector projection technique. In this paper we present a novel efficient bit-vector projection technique, for sparse and dense datasets. To check the efficiency of our bit-vector projection technique, we present a new frequent itemset mining algorithm Ramp (Real Algorithm for Mining Patterns) build upon our bit-vector projection technique. The performance of the Ramp is compared with the current best (all, maximal and closed) frequent itemset mining algorithms on benchmark datasets. Different experimental results on sparse and dense datasets show that mining frequent itemset using Ramp is faster than the current best algorithms, which show the effectiveness of our bit-vector projection idea. We also present a new local maximal frequent itemsets propagation and maximal itemset superset checking approach FastLMFI, build upon our PBR bit-vector projection technique. Our different computational experiments suggest that itemset maximality checking using FastLMFI is fast and efficient than a previous will known progressive focusing approach.

## 1 INTRODUCTION

Since the introduction of association rules mining (**ARM**) by Agrawal et al. [1], it has now became one of the main pillars of data mining and knowledge discovery tasks and has been successfully applied in sequential pattern mining, emerging pattern mining, multidimensional pattern mining, classification, maximal and closed itemset mining [3], [9], [16], [21], [4]. Using the support-confidence framework presented by [2], the problem of mining the complete association rules from transactional dataset is divided into two parts – (a) finding complete frequent itemsets with support (an itemset's occurrence in the dataset) greater than minimum support threshold, (b) generating association rules from frequent itemsets with confidence greater than minimum confidence threshold. In practice, the first phase is the most time-consuming task, which requires the heaviest frequency counting operation for each candidate itemset.

Let *TDS* be our transactional dataset and *I* be a set of distinct items in the *TDS*. Each transaction *t* in the *TDS* consist a subset of items such as $t_I$ Í *I*. We call *X* an itemset if it fulfill of *X* Í *I*. Let *min-sup* be our minimum support threshold, we call an itemset *X* frequent if its support *(support(X))* is greater than *min-sup*; otherwise infrequent. By following the Apriori property [2] an itemset *X* cannot be a frequent itemset, if one of its subset is infrequent. We denote the set of all frequent itemset by **FI**. If *X* is frequent and no superset of *X* is frequent, we say that *X* is a maximal frequent itemset, the set of all maximal frequent itemsets is denoted by **MFI**. If *X* is frequent and no superset of *X* is as frequent as *X*, we say that *X* is a closed frequent itemset; similarly the set of all closed frequent itemset is denoted by **FCI**. Thus the following condition is straight-forward holds: **MFI** ⊆ **FCI** ⊆ **FI**.

When the frequent patterns are long (more than 15 to 20 items), mining **FI** becomes infeasible because of the exponential number of frequent itemsets. For example, Apriori [2] a well known frequent itemset mining algorithm considers all the $2^k$ subsets as candidate itemsets, where $k$ is the cardinality of itemset $X$. Thus algorithms for mining **FCI** [21], [27] have been proposed, since **FCI** is enough to generate association rues. However, **FCI** could also be exponentially large as the **FI**. As a results, researchers turn to find **MFI**. Given the set of **MFI**, it is easy to analyze many interesting properties of the dataset, such as the longest patterns, the overlap of the **MFI** etc. Secondly all **FI** can be built up from **MFI** and can be counted for support in a single dataset scan. Moreover, we can focus on part of the **MFI** to do supervised data mining.

In last 5 years, a number of algorithms have been proposed for efficient enumeration of **FI**, **MFI** and **FCI** [15], [8], [21], [27]. Considering the importance of frequent itemset mining, in 2003 and 2004 two international conference were organized for only frequent itemset mining implementations (FIMI2003 and FIMI2004) [11], where number of authors presented their ideas and original implementations. According to the **FIMI** report [11], itemset frequency calculation is considered to be a most important factor in overall frequent itemset mining, which highly depends upon dataset representation approach and dataset projection/compression technique. In practice FP-Tree [15], array based layout approach [22] and vertical bit-vector representation approach [8] are three main and widely used dataset representation approaches. Most of efficient frequent itemset mining algorithms e.g. AFOPT [18], FPgrowth-(zhu) [14] and MAFIA [8] use these dataset representation approaches for frequent itemset mining. According to the **FIMI** report [11], currently there is no global best algorithm exists, that is faster on all types of datasets e.g. sparse as well as dense. As indicated in **FIMI** report [11] and [18], MAFIA [8] a maximal frequent itemset algorithm is faster and efficient for dense type datasets, while AFOPT [18] and FPgrowht-(zhu) [14] are faster and efficient algorithms for sparse type datasets.

## 1.1 Contribution of this paper

As we have mentioned earlier, MAFIA a maximal itemset mining algorithm [8] (using bit-vector representation approach) is considered to be most efficient algorithm for dense datasets mining. The basic strategy of MAFIA is that, it traverses search space in depth first search order, and on each node of search space it removes infrequent item from its node's tail using dynamic reordering. To check the frequency (support) of itemset it performs a bitwise-AND (bitwise-$\wedge$) operation on head and tail item bit-vectors. Calculating frequency using bit-vectors representational approach is efficient when the dataset is dense, but highly inefficient when the items bit-vectors contain more zeros than ones, resulting in many useless counting operations, which usually happens in the case of sparse datasets. To handle the bit-vectors sparseness problem, MAFIA proposed a bit-vector projection technique known as projected bitmap. The main deficiency of projection using projected bitmap technique is that, it requires a high processing cost (time) for its construction. Due to this reason, MAFIA uses adaptive compression [8], since projection is done only when saving from the compressed bit-vectors outweigh the cost of projection. However, with adaptive compression, projection cannot be possible on all nodes of search space.

In this paper we present a novel bit-vector projection technique, we call it Projected-Bit-Regions (**PBR**) bit-vector projection technique. The main advantages of projection using PBR are that – (a) it consumes a very small processing cost and memory space for projection and (b) it can be easily apply able on all nodes of search space without requiring any adaptive approach. To check the efficiency of our bit-vector projection idea, we present a new itemset mining algorithm (Ramp) build upon **PBR** bit-vector projection technique for all, maximal and closed frequent itemset mining. We also present some efficient implementation techniques of Ramp, which we experienced in our implementations. An earlier version of this paper is accepted in the *10$^{th}$ Pacific Asia Conference on Knowledge and Data Discovery 2006* [5]. In this paper we enhance the Ramp algorithm for maximal and closed itemset mining. Our different experimental results on dense and sparse benchmark datasets suggest that the Ramp is faster than the current best algorithms which marked good scores on FIMI03 and FIMI04 [11]: Fpgrowth-(zhu) [14], AFOPT [18] and MAFIA [8]. This shows the effectiveness of our **PBR** bit-vector projection technique and implementation ideas.

## 2 RELATED WORK

Consider an example dataset of store sales shown in Figure 1. There are 17 different items (A., B, C, D, E,

F, G, H, I, J, K, L, M, N, O, P, Q) and the dataset consists of seven customers who bought these different items. Figure 2 shows all the frequent itemsets that occurr in at least two transactions, i.e., *min_sup = 2.*

| **Transaction** | **I** |
|---|---|
| 01 | A, B, C, F, G, L |
| 02 | A, B, ,H, I |
| 03 | B, E, J, O |
| 04 | C, E, M, I, Q |
| 05 | A, B, D, N |
| 06 | A, B, C, D, K |
| 07 | A, P |

**I** = Items of each transaction

Figure 1: A sample transactional dataset representation.

| **Support** | **FI** |
|---|---|
| 71%(5) | A, B |
| 57% (4) | AB |
| 42% (3) | C |
| 28% (2) | D, E, I, AC, AD, BC, BD, ABC, ABD |

**FI** = Frequent Itemsets

Figure 2: Frequent itemsets of Figure 1 dataset with min_sup = 2.

### 2.1 Frequent Itemset Mining Using Apriori Algorithm

The Apriori algorithm by Agrawal et al. [2] is considered as one of most well known algorithm in frequent itemset mining problem. Apriori uses a complete, bottom-up search, with a horizontal layout and prune infrequent itemsets using anti-monotone Apriori heuristic: if any length *k* pattern is not frequent in the database, its length *(k+1)* super pattern can never be frequent. Apriori is an iterative algorithm that counts itemsets of specific length in a given database pass. The algorithm starts with scanning all transactions of the transactional dataset and computes frequent items. All those items which have frequency (support) lower than *min_sup* are pruned from search space at root node. For example in Figure 1, items á*A: 5, B: 5, C: 3, D: 2, E: 2, I: 2*ñ are only locally frequent items. Next, a set of frequent candidate 2-itemsets are formed from the frequent items i.e. á*AB, AC, AD, AE, AI, BC, BD, BE, BI, EI*ñ. Another dataset scan collects their supports and removes all those 2-itemsets which have frequency lower than *min_sup*. For example, candidate 2-itemsets á*AE, AI, BE, BI, EI*ñ have frequency less than *min_sup,* so these are removed from frequent 2-itemsets collection. Then these frequent 2-itemsets are retained and used for frequent candidate 3-itemsets. This process is repeated until all frequent itemsets have been discovered. There are three main steps in the algorithm:

1 Generating frequent candidate (k+1)-itemsets, by joining the frequent itemsets of previous pass k.

2 Deleting those subsets which are infrequent in the previous pass *k* without considering the transactions in the dataset.

3 Scanning all transactions to obtain frequent candidate (k+1)-itemsets.

Although, Apriori presented by Agrawal is very effective method for enumerating frequent itemsets of sparse datasets on large support threshold, but the basic algorithm of Apriori encounters some difficulties and takes large processing time on low support threshold. We list here some main deficiencies of Apriori that make it an unattractive solution for mining frequent itemsets.

1. Apriori encounters difficulty in mining long pattern, especially for dense datasets. For example, to find a frequent itemsets of $X = \{1\ldots200\}$ items. Apriori has to generate-and-test all $2^{200}$ candidates.
2. Apriori algorithm is considered to be unsuitable for handling frequency counting, which is considered to be most expensive task in frequent itemsets mining. Since Apriori is a level-wise candidate-generate-and-test algorithm, therefore it has to scan the dataset *200* times to find a frequent itemsets $X = X_1\ldots X_{200}$.
3. Even though Apriori prunes the search space by removing all *k* itemsets, which are infrequent before generating candidate frequent (*k+1*)-itemsets, it still needs a dataset scan, to determine which candidate (*k+1*) itemsets are frequent and which are infrequent. Even for datasets which have *200* items, determining *k*-frequent itemsets by repeated scanning the dataset with pattern matching takes a large processing time.

## 2.3 Frequent Itemset Mining with Pattern-Growth Approach

Apriori and its variants enumerate all frequent itemsets by repeatedly scanning the dataset and checking the frequency of candidate frequent *k* itemsets by pattern matching. This whole process is costly especially if the dataset is dense and has long patterns, or a low minimum support threshold is given. To increase the efficiency of frequent itemset mining Han et al. [15] presented a novel method called pattern-growth. The major advantage of frequent itemsets mining using pattern-growth approach is that it removes the costly operation of repeatedly scanning the dataset in each iteration and generating and testing infrequent candidate itemsets. In simple words pattern growth removes the costly candidate-generate-and-test operation of Apriori type algorithms. The method requires only two dataset scans for counting all frequent itemsets. The first dataset scan collects the support of all frequent items. The second dataset scan builds a compact data structure called FP-tree. Each node of FP-tree corresponds to an item which was found frequent in first dataset scan. Next, all frequent itemsets are mined directly from this FP-tree without concerning the transactional dataset.

The pattern-growth approach mines all frequent itemsets on the basis of FP-tree. The main strategy of pattern-growth approach is that it traverses search space in depth first order, and on each node of search space it mines frequent itemsets on the basis of conditional patterns and creates child FP-tree (also called projected database).

## 2.4 Improvements over Pattern-Growth Approach

Mining frequent itemsets with pattern-growth approach by recursive creation of conditional patterns and conditional projected databases (FP-tree) is highly efficient and space effective for dense datasets, if lot of transactions shares same prefix. However, if the dataset is huge and sparse then recursive creation of projected FP-tree requires a huge space, which is almost equal to the total number of frequent itemsets. In past few years several improvements over basic pattern-growth [15] have been proposed.

### *2.4.1 Pseudo-construction Techniques*

Constructing conditional patterns and conditional databases (FP-trees) by physical representation approach takes a huge space, especially for sparse datasets on low support threshold. As an alternate, recently some pseudo construction techniques [22], [17] have been proposed, i.e. using pointers pointing to transactions in upper level of conditional databases. However, pseudo-construction cannot reduce traversal cost as effectively as physical construction.

#### *2.4.2 AFOPT: Ascending Frequency Order Prefix Trees*

AFOPT proposed by [18] is considered to be as one of the major improvement over pattern-growth approach. The main strategy of AFOPT is that it traverses conditional databases (FP-tree) top-down traversal rather than bottom-up approach, and construct conditional databases (FP-trees) by ascending frequency ordering method. The top-down traversal strategy is capable of minimizing the traversal cost of a conditional database, where the ascending frequency order method is capable of minimizing the total number of conditional databases. On first dataset scan, all infrequent items which have support less than the *min_sup* are removed from frequent items list. In second dataset scan, the AFOPT-tree constructs, all nodes which have single branch are stored with array which save the space and construction cost. Next the itemset mining process from root node is started. To enumerate a frequent itemsets from *t's* conditional database, AFOPT mines it with the following three steps.

1. In first step, $D_t$ is traversed by top-down, and the support of all items in $D_t$ is counted.
2. In second step, a new frequent $D_t$ is constructed which include only the frequent extension of first step i.e. frequent items.
3. In third step, the sub trees of *t* are merged with *t's* right siblings. The merging operation involves only pointer adjustments and support updates. The main aim of this step is that, it decreases the size of trees.

#### *2.4.3 FPgrowth-(zhu): Efficient Implementation of Pattern Growth Approach*

Different experiments presented by Grahne et al. [14] demonstrate that almost 80% of frequent itemsets mining time using pattern-growth approach is spend on traversing *t*'s conditional FP-tree and counting support of items in *t*'s conditional FP-tree. In general, to mine frequent itemsets from *t*'s conditional FP-tree two steps are involve – (a) counting support of items in *t* (b) constructing conditional tree from *t* conditional database. To increase the efficiency of pattern-growth approach they present an array-based data structure called FP-array which is considered to be an efficient implementation of pattern growth approach. With FP-array the algorithm needs to scan each FP-tree only once for each recursive call to enumerate all frequent itemsets.

The main strategy of mining frequent itemsets by FP-array technique is that for each frequent item *i* in conditional database *t*, the algorithm constructs a conditional FP-tree and FP-array. The FP-array is a two dimensional array of size *m*m*, where *m* is total number of items in FP-tree. The main illustration behind creating array structure for each conditional database is that, it contains the support of all items in the conditional database (FP-tree) and can be accessed in O (1) cost, which eliminates the costly first step of normal pattern growth approach. FP-array of conditional database can be constructed at the same time, when its conditional database (FP-tree) is constructed from parent conditional database.

## 2.5 MAFIA: Maximal Frequent Itemset Algorithm using Vertical Bit-vector Representation Approach

Burdick et al. [8] presented a simple hardware efficient dataset representation scheme (vertical bit-vector) and a frequent itemset mining algorithm called MAFIA. To count the frequency of any itemset, the algorithm performs a simple bitwise-AND operation on two bit-vectors and resulting ones represents the frequency of that itemset. On 32-bit CPU the algorithms performs a bitwise-AND operation on 32-bits per operation. As indicated in [11], [18] Mafia is considered to be one of the fastest algorithm for small dense datasets mining.

Mafia also has parent equivalence pruning (**PEP**) and differentiates superset pruning into two classes **FHUT** and **HUTMFI**. For a given node *X:aY*, the idea of **PEP** is that if *sup(X)=sup(Xa)*, i.e. every transaction containing *X* also contains the item ***a***, then the node can simply be replaced by *Xa:Y*. The **FHUT** uses leftmost tree to prune its sister, i.e., if the entire tree with root *Xa:Y* is frequent, then we do not need to explore the sisters of the node *Xa:Y*. The **HUTMFI** uses to use the known **MFI** set to prune a node, i.e., if itemset of *XaY* is subsumed by some itemset in the **MFI** set, the node *Xa:Y* can be pruned. MAFIA also uses dynamic reordering to reduce the search space. The results show that **PEP** has the biggest effect of the above pruning methods (**PEP**, **FHUT**, and **HUTMFI**).

| Transaction | A | B | C | D | E | I |
|---|---|---|---|---|---|---|
| 01 | 1 | 1 | 1 | 0 | 0 | 0 |
| 02 | 1 | 1 | 0 | 0 | 0 | 1 |
| 03 | 0 | 1 | 0 | 0 | 1 | 0 |
| 04 | 0 | 0 | 1 | 0 | 1 | 1 |
| 05 | 1 | 1 | 0 | 1 | 0 | 0 |
| 06 | 1 | 1 | 1 | 1 | 0 | 0 |
| 07 | 1 | 0 | 0 | 0 | 0 | 0 |

Figure 3: Vertical bit-vector representation of Figure 1 dataset.

# 3 Mining Frequent Itemset Using Vertical Bit-vector Representation Approach

In a vertical bitmap, there is one bit for each transaction of dataset. If item *i* appear in maximal pattern *j*, then the bit of *j* of the bitmap of item *i* is set to one; otherwise the bit is set to zero. Figure 3 shows the vertical bit-vector representation of Figure 1 dataset. To count the frequency of an itemset e.g. *{AB}* we need to perform a bitwise-AND operation on bit-vector *{A}* and bit-vector *{B}*, and resulting ones in bit-vector *{AB}* will be the frequency of itemset *{AB}*.

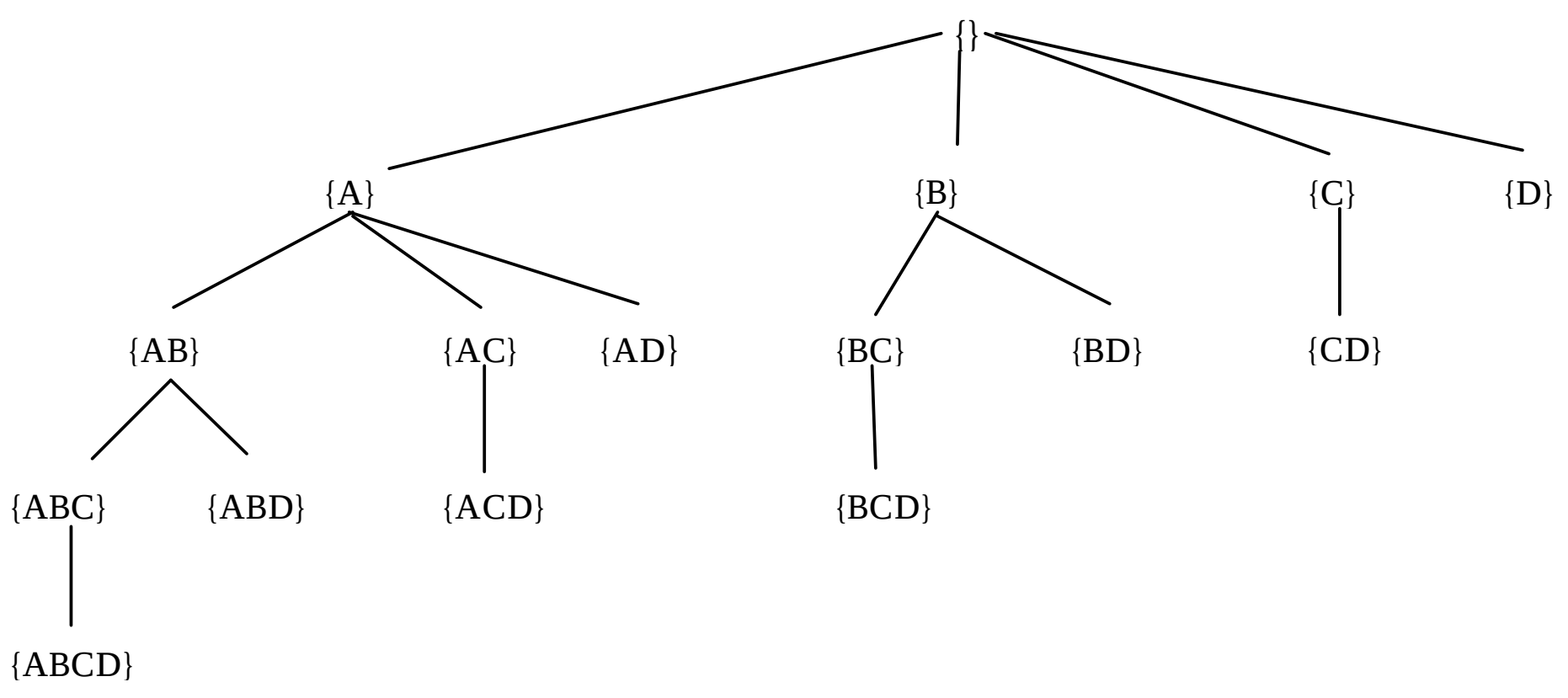

Figure 4: Lexicographic tree of four items ⟨A, B, C, D⟩.

## 3.1 Itemset Generation using Bit-vector Representation Approach

Let < be some lexicographical order of the items in *TDB* such that for every two items *a* and *b, a ¹ b: a < b or a > b*. The search space of frequent itemset mining can be considered as a lexicographical order [25], where root node contains an empty itemset, and each lower level *k* contains all the *k*- itemsets. Each node of search space is composed of head and tail elements. Head denotes the itemset of node, and items of tail are possible extensions of new child itemsets. For example with four items *{A, B, C, D},* in Figure 4 root's head is empty *á()ñ* and tail is composed with all of items *á(A, B, C, D)ñ*, which generates four possible child nodes *{head á(A)ñ: tail á(BCD)ñ}, {head á(B)ñ: tail á(CD)ñ}, {head á(C)ñ: tail á(D)ñ}, {head á(D)ñ: tail á{}ñ}.*

This itemset search space can be traverse either by depth first order or breadth first order. At each node of search space, infrequent items from tail are removed by dynamic reordering and ordered by increasing support which keeps the search space as small as possible.

## 3.2 Itemset Frequency Calculation

To check, whether every item *X* in tail *n* is frequent or infrequent, we must check its frequency (support) in TDB. Calculating itemset frequency in bit-vector representation requires applying bitwise-$\wedge$ operation on *n.head* and *X* bit-vectors, which can be implemented by a loop; which we call **simple loop**, where each iteration of **simple loop** apply bitwise-$\wedge$ operation on some region of *n.head* with *X* bit-vectors. Since 32-bit CPU supports 32-bit $\wedge$ per operation, hence each region of *X* bit-vector is composed of 32bits (represents 32 transactions). Therefore calculating frequency of each itemset by using **simple loop** requires applying bitwise-$\wedge$ on all regions of n.head with *X* bit-vectors. However, when the dataset is sparse, and each item is presented in few transactions, then counting itemset frequency by using **simple loop** and applying bitwise-$\wedge$ on those regions of bit-vectors which contain zero involves many unnecessary counting operations. Since the regions which contain zero, will contribute nothing to the frequency of any itemset, which will be superset of *n* itemset. Thereby removing these regions from head bit-vectors in earlier stages of search space is more beneficial and useful.

### 3.3 Bit-vector Projection using Projected Bitmap

The major weakness of bit-vector representation is its sparseness on sparse datasets especially on very low support thresholds. Since each frequent item is presented in very limit number of transactions, therefore its bit-vector contains more zero number regions than non-zero number regions, since both the absence and presence of the item in a transaction need to be represented. Given the frequency counting mechanism presented in section 3.2, applying bitwise $\wedge$ operation on those bit-vector regions of $X$ with $Y$ which contains zeroes involves many unnecessary counting operations.

Burdick et al. [8] present a projected bitmap projection technique to increase the efficiency of bit-vector representation on sparse datasets. The main idea of their bit-vector projection technique is that, the transaction $T$ which does not contain the itemset $X$ (*X's bit-vector has a 0 in bit T*) will not provide useful information for counting any itemset support which will be superset of $X$, and thus it can be ignored in all subsequent operations. So, in their bit-vector projected bitmap projection technique, they copied all bits of $X$ which contain a value of 1, with all $X$ tail's items bits, to another bit-vector memory which they call projected-bit-vector. In the projected-bit-vector for $X$ all positions have value 1, since each transaction corresponds to a transaction needed for computation *X È Y* or any itemset which is superset of $X$.

Note however that though the bit-vector projection idea presented by [8] is very efficient for dense datasets, but copying item $X$ bit positions at node n with all of its tail items bits to another memory space at each node of search is inefficient and takes a huge processing time for construction of projected bitmap. Sometime this construction cost is so high that the total mining time becomes worse than the time with no bitmap projection. To further increase the efficiency of projected bitmap idea, in [8] they present a projected bitmap adaptive projection technique which uses a rebuilding threshold for projection. Since, with rebuilding threshold projection is done only when its saving is greater than the cost of projection construction. Therefore, with adaptive compression, projection cannot be possible on all nodes of search space.

# 4 BIT-VECTOR PROJECTION USING PROJECTED-BIT- REGIONS (PBR) APPROACH

For efficient projection of bit-vectors, the goal of projection should be such as, to bitwise-$\wedge$ only those regions of head bit-vector 〈*bitmap(head)*ñ with tail item *X* bit-vector 〈*bitmap(X)*ñ which contains a value greater than zero and skip all others. Obviously for doing this, our counting procedure must be so powerful and have some information which guides it, that which regions are important and which ones it can skip. To achieve this goal, we propose a novel bit-vector projection technique **PBR** (Projected-Bit-Regions). With projection using **PBR**, each node *Y* of search space contains an array of valid region indexes $PBR_{\text{áYñ}}$ which guides the frequency counting procedure to traverse only those regions which contain an index in array and skip all others. Figure 5 shows the code of itemset frequency calculation using **PBR** technique. In Figure 5, line 1 first retrieves a valid region index $\ell$ in 〈*bitmap (head)*ñ and line 2 apply a bitwise-$\wedge$ on 〈*bitmap (head)*ñ with 〈*bitmap (X)*ñ on region $\ell$.

(1) **for** each region index $\ell$ in projected-bit-regions of head

(2) AND-result = bitmap-X [$\ell$] $\wedge$ bitmap-head [$\ell$]

(3) Support-X = Support-X + number of ones(AND-result)

Figure 5: Itemset frequency calculation using Projected-Bit-Regions (PBR).

| **Transaction** | **I** | **FI** |
|---|---|---|
| 01 | A B C F G L | A B C |
| 02 | A B H I | A B I |
| 03 | B E J O | B E |
| 04 | C E M I Q | C E I |
| 05 | A B D N | A B D |
| 06 | A B C D K | A B C D |
| 07 | A P | A |

**I** = Items of transactions
**FI** = Frequent items

Figure 6: A sample transactional dataset with frequent items.

One main advantage of bit-vector projection using **PBR** is that, it consumes a very small processing cost for its creation, and hence can be easily applied on all nodes of search space. At any node, projection of child nodes can be created either at the time of frequency calculation if pure depth first search is used, or at the time of creating head bit-vector if dynamic reordering is used. The strategy of creating $PBR_{\text{áXñ}}$ at node ***n*** for each tail item *X* is that, when the **PBR** of ⟨*bitmap(n)ñ* are bitwise-∧ with ⟨*bitmap(X)ñ* a simple check is perform on each bitwise-∧ result. If the value of result is greater than zero, then an index is allocated in $PBR_{\text{án.head È Xñ}}$. The set of all indexes which contain a value greater than zero makes the projection of *{head È X}* node.

## 4.1 Memory Requirements

Some other advantages of projection using **PBR** are that, it is a very scalable approach and consumes a very small amount of memory during projection and can be applicable on very large sparse datasets. Scalability is achieved as, we know by traversing search space in depth first order; a single tree path is explored at any time. Therefore a single **PBR** array for each level of path needs to remain in memory. As a preprocessing step a **PBR** array for each level of path is created and cached in memory. At itemset generation time different paths of search space (tree) can share this memory and do not need to create any extra projection memory during itemset mining.

## 4.2 Projection Example

Let the second column of Figure 6 shows the transactions of a sample *TDB*. Let the minimum threshold be min_sup = 2. For ease of explanation let us make two assumptions – (a) each region of bit vectors contain only a single bit (b) in dynamic reordering infrequent tail items are removed from nodes, but a static alphabetical order is followed instead of ascending frequency order.

In first dataset scan, frequency of each item in *TDB* is calculated and all those items which have frequency less than min_sup are removed from *TDB* transactions. In Figure 6 column 3 shows the transactions contained only the frequent items. After first dataset scan, the following items are found frequent *á(A: 5, B: 5, C: 3, D: 2, E: 2, I: 2)ñ* where notation *A: 5* means item *A* has frequency 5. Next a bit-vector of each frequent item is created, where each bit-vector consist seven regions.

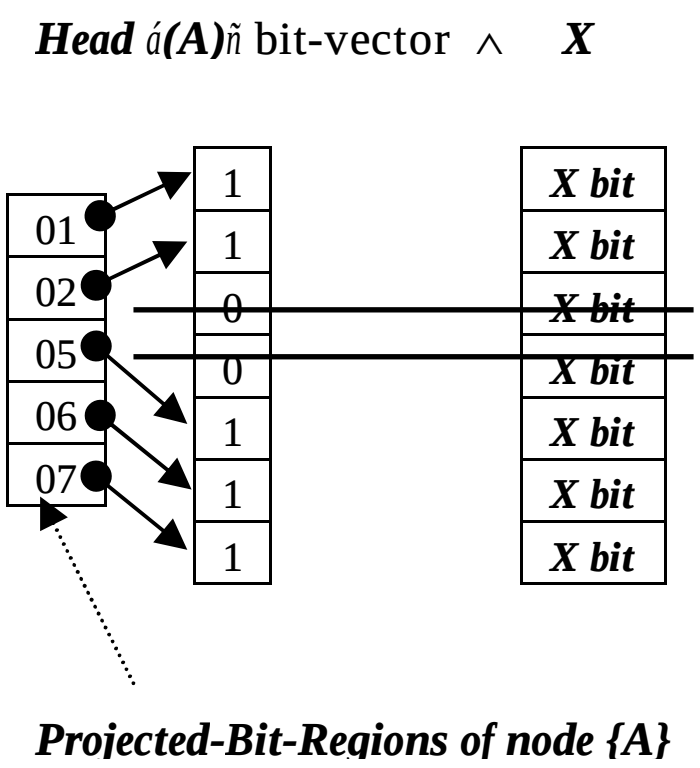

Figure 7: Itemset *(A È X)* frequency calculation using **PBR**.

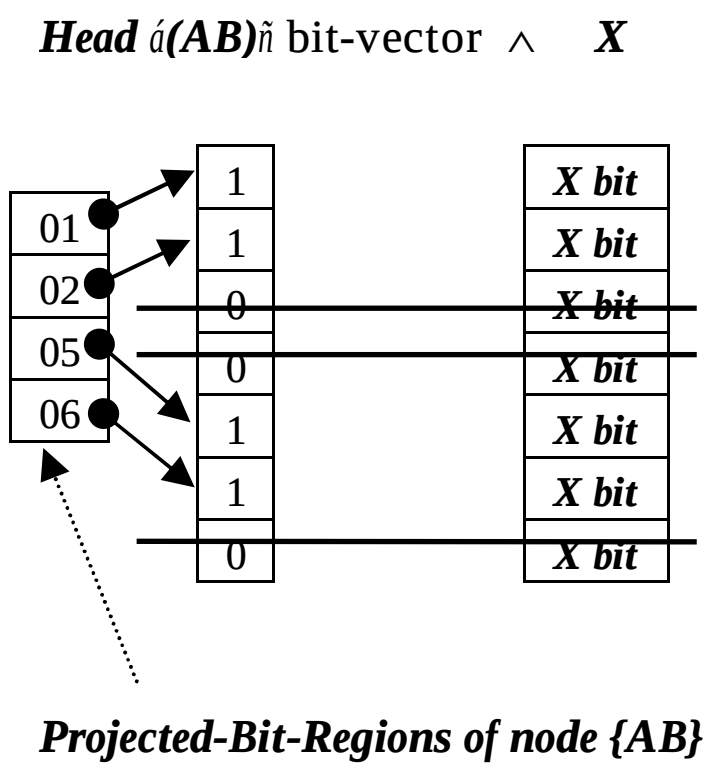

Figure 8: Itemset *(AB È X)* frequency calculation using **PBR**.

Afterwards, itemset mining process from root node is started. At root node head itemset is *{}*, therefore the PBR of each tail item *X = á(A, B, C, D, E, I)ñ* is created by traversing all *ábitmap(X)ñ* regions and indexing only those which contain a value greater than 0. For example *ábitmap (A)ñ* contains a value greater than zero in bit regions *á(01, 02, 05, 06, 07)ñ*; thereby these are its PBR *indexes*. Since we are traversing search space in depth first order and item *{A}* is first in alphabetical order. Therefore root node creates a child node with *head á(A)ñ* and *tail á(BCDEI)ñ* and iterates to *node {A}*.

At *node {A}*, frequency of each tail item *X* is calculated by applying bitwise-∧ on *ábitmap (A)ñ* with *ábitmap(X)ñ* on $PBR_{\text{áAñ}}$. Figure 7 shows the frequency calculation process of itemset *á(X È head.{A})ñ* by using PBR. After frequency calculation tail items *á(B: 4, C: 2, D: 2)ñ* are found locally frequent at *node {A}*. Since item *{B}* in *tail.{A}* is next in alphabetical order, therefore the *node {A}* creates a new child node with *head á(AB)ñ* and *tail á(C, D)ñ*. *PBR = á(01, 02, 05, 06)ñ* of child *node {AB}* at *node {A}* are created by applying bitwise-∧ on *á(bitmap (A)ñ* with *á(bitmap (B)ñ* on $PBR_{\text{áAñ}}$. The mining process then moves to *node*

*{AB}* with *head á(AB)ñ* and *tail á(CD)ñ*.

Similarly at *node {AB}*, frequency of each tail item *X* is calculated by applying bitwise-∧ on *ábitmap (AB)ñ* with *ábitmap(X)ñ* on $PBR_{áABñ}$. Figure 8 is showing frequency calculation process at *node {AB}*. After frequency calculation tail items *á(C: 2 D: 2)ñ* are found locally frequent at *node {AB}*. Since item *{C}* in *tail.{AB}* is next in alphabetical order, therefore the mining process creates a child node with *head á(ABC)ñ* and *tail á(D)ñ*. *PBR = á(01, 06)ñ* of child *node {ABC}* at *node {AB} are* created by applying bitwise-∧ on *ábitmap (AB)ñ* with *á(bitmap (C)ñ* on $PBR_{áABñ}$. Afterward, the mining process iterates to *node {ABC}* with *head á(ABC)ñ* and *tail á(D)ñ*.

At *node {ABC}*, all tail items are infrequent; therefore the mining process stops and backtracks at *node {AB}* and mine other items of its *tail á(D)ñ*. Since *item {D}* is next in alphabetical order, so the mining process creates a new child *node {ABD}* with *head á(ABD)ñ* and *tail á({})ñ*. At *node {ABD}*, tail is empty so the mining process stops and backtracks at *node {A}*. Similarly, the mining process mines other items of *node {A}* tail in same fashion. After mining all tail items of *node {A}* tail, the mining process then backtracks at root node, and mines other tail items of root in depth first order. Due to the lack of space we could not give a detail example, but the basic idea for mining other itemsets is similar.

# 5 RAMP: REAL ALGORITHM FOR MINING PATTERNS

## 5.1 RAMP-ALL: All Frequent Itemset Mining

The basic strategy of *Ramp* for mining all frequent itemset is that, it traverses search space in depth first order. At any node *n*, infrequent items from tail are removed by dynamic reordering and new node *m* for every tail item *X* in tail *n*, is generated such as *m.head = n È X* and *m.tail = n.tail – X*. Items in *m.tail* are reordered by increasing support which keeps the search space as small as possible. For frequency counting, item *X* bit-vector is bitwise-∧ with *n.head* bit-vector on $PBR_{ánñ}$. The pseudo code of *Ramp* is described in Figure 9.

```
Ramp-all (Node n)

(1)  for each item X in n.tail
(2)     for each region index ℓ in PBR⟨n⟩
(3)              AND-result = bit-vector[ℓ] ∧ head-
                 bit-vector of n [ℓ]
(4)              Support[X] = Support[X] + number of
                 ones(AND-result)

(5)  Remove infrequent items from n.tail, and reorder
     them by increasing support

(6)  for each item X in n.tail
(7)     m.head = n.head È X
(8)     m.tail = n.tail – X
(9)      for each region index ℓ in PBR⟨n⟩
            AND-result = bit-vector[ℓ] ∧ head-bit-
            vector[ℓ]
(10)         If AND-result > 0
(11)            Insert ℓ in PBR⟨m⟩
(12)            head bit-vector of m [ℓ] = AND- result

(13)     Ramp-all (m)
```

Figure 9: Ramp-all: Pseudo code for mining all frequent itemsets using PBR projection technique.

## 5.2 Efficient Implementation Techniques

In this section, we provide some efficient implementations ideas, which we experienced in our Ramp implementations. Some of our implementations ideas such as **2-Itemset-Pair** (section 5.2.3) and **Fast-Output-FI** (section 5.2.4) are generic, which can be apply on any type of frequent itemset algorithm.

### *5.2.1 Eliminating Redundant Frequent Counting Operations (ERFCO)*

The code which we describe in the Figure 9 performs exactly two frequency counting operations for each frequent tail item *X* at any node *n* of search space. First, at the time of performing dynamic reordering at node *n* and second, for creating *{X È n.head}* bit-vector. The itemset frequency calculation process which is considered to be most expensive task (penalty) in overall itemset mining [11], the bit-vector representation approach suffers this penalty twice for each frequent itemset. The second counting operation which we can say is redundant, occurs due to gain efficiency in 32bit CPU and can be eliminated with some efficient implementation, which we describe below.

In *Ramp*, at the start of algorithm two large heaps, one for head bit-vectors and one for PBR are created (with 32bit per heap slot size). Next, at any itemset *X* frequency calculation time a simple check is performed to ensure that is there sufficient space left in both heaps. If the response is “yes” then the head bit-vector of *X* and $PBR_{\langle X \rangle}$ are created at the same time when its frequency is calculated, otherwise normal procedure is followed. The main difference is that, with the efficient implementation bitwise-$\wedge$ results and regions indexes are written in heaps instead of tree path levels memories. The size of heaps should be so enough that it can store any frequent item subtree. From our implementation point of view, we suggest that heap size double the total number of transactions is enough for very large sparse datasets. In our *Ramp* implementation it completely eliminated the second frequency counting operation while requiring very little amount of memory.

### *5.2.2 Increasing Projected-Bit-Regions Density (IPBRD)*

The bit-vector projection technique described in section 4 does not provide any compaction or compression mechanism to increase the density in bit-vector regions. As a result, on the sparse dataset only one or two bits are set in each bit-vector region, which not only increase the projection length but also it is not possible to achieve true 32bit CPU performance. To increase the density in bit-vector regions the *Ramp* starts with an array-list [22]. Next at root node, a bit-vector representation for each frequent item is created which provide a sufficient compression and compaction in bit-vectors regions. Sufficient improvements are obtained in *Ramp* by using this approach.

### *5.2.3 2-Itemset Pair*

There are two methods to check whether current itemset is frequent or infrequent. First, to directly compute its frequency from *TDB*. Second one, which is efficient known as 2-Itemset pair. If any 2-Itemset pair of any itemset is found infrequent, then by following Apriori [2] property itemset is consider to be as infrequent. In AIM [10] almost the same approach was used with the name *efficient initialization*. However AIM used this approach only for those itemsets which contain a length equal to two. In *Ramp* we extend the basic approach and apply 2-Itemset pair approach also on those itemsets which contain a length more than two. We know any itemset which contains a length more than two, is the superset of its entire 2-Itemset pairs. Before counting its frequency from transactional dataset, *Ramp* checks its 2-Itemset pairs. If any pair is found infrequent then that itemset is automatically considered to be infrequent.

### *5.2.4 Writing Frequent Itemsets to Output File (Fast-Output-FI)*

When the dataset is dense and contains millions of frequent itemsets on low support threshold, almost 90% of overall mining time is spent on writing frequent itemsets to output file [24]. We have noted hat some of previous implementations e.g. AFOPT [19], PatriciaMine [22], fpgrowth-zhu [14] write output itemsets one by one, which increases the context switch and disk rotation times and degrades their algorithm performance. A better approach which we use in *Ramp* is to write itemsets to output file only when a sufficient number of itemsets are mined in memory. In *Ramp* we find that, writing itemsets using this approach sufficiently decreases the processing time of algorithm. Fast rendering of integers to strings is also an important factor for dense datasets. Since all the frequent itemsets are mined in the form of integers, while output file is written in the form of text. A fast rendering procedure of converting integers to strings can also improve the performance of algorithm.

## 6 MINING MAXIMAL FREQUENT ITEMSETS

Mining **MFI** is considered to be more advantage able than mining **FI**, since it mines small and useful long patterns. However, mining **MFI** is more complicated than mining **FI**, since for each candidate maximal itemset; we not only check its frequency (support) but also its maximality, which takes O (**MFI**) cost is worst case. In practice, checking itemset maximally is considered to be an important factor in **MFI** mining. As per literature review two techniques, progressive focusing [12] and MFI-tree [13] has been proposed for checking **MFI** maximally, efficiently. Where, progressive focusing is widely used in most of the **MFI** mining algorithms [8], [18].

In our Ramp-max algorithm, we proposed another technique for itemset maximality checking name as FastLMFI using bit-vector representation approach and our **PBR** projection technique (section 3). In our extensive experiments we found that checking itemset maximality using FastLMFI is more fast and efficient than previous progressive focusing approach. An earlier version of our FastLMFI is presented in IEEE-AICCSA 2006 [6], since then we proposed several ideas to improve it.

### 6.1 Local Maximal Frequent Itemsets

Let *list (MFI)* be our currently known maximal itemsets, and let *Y* be our new candidate maximal itemset. To check if *Y* is subset of any known mined maximal itemset, we need to performed a maximal superset checking, which takes *O(MFI)* in worst case. To speedup the superset checking cost, local maximal frequent itemset (**LMFI**) has been proposed. **LMFI** is a divide and conquer strategy, which contains only those relevant maximal itemsets, in which *Y* appears as a prefix.

Any maximal itemset pattern containing *P* itemsets can be a superset of $P\grave{E}_{subsets(P)}$ or $P\grave{E}_{freq_ext(P)}$. The set of $P\grave{E}_{freq_ext(P)}$ is called the local maximal frequent itemset with respect to *P*, denoted as $LMFI_p$. To check whether *P* is a subset of some existing maximal frequent itemsets, we only need to check them against $LMFI_p$, which takes O($LMFI_p$) cost. If $LMFI_p$ is empty, then *P* will be our new maximal itemset, otherwise it is subset of $LMFI_p$.

### 6.2 FastLMFI: Local Maximal Frequent Itemsets Propagation and Itemset Maximality Checking

In this section we explain the **LMFI** propagation and **MFI** superset checking using bit-vector representation approach and our **PBR** projection technique. From implementation point of view, progressive focusing $LMFI_p$ (where *P* is any node) can be constructed either from its parent $LMFI_p$ or sibling of *P*. With progressive focusing, construction of child *LMFIs* takes two steps [12]. First, project them in parent $LMFI_{p+1}$. Second, pushing and placing them in top or bottom of *list (MFI)* for constructing $LMFI_{p+1} = LMFI_{p\,\grave{E}\,\{i\}}$, where *i* is tail item of node *P*.

We here list up the some advantages of our FastLMFI over progressive focusing approach.

1. Creating child $LMFI_{p+1}$ in one step, rather than into two steps. By using our **PBR** bit-vector projection technique, we can completely eliminate second step. It may be noted the second step is more costly (removing and adding pointers) than first step.
2. Optimizing first step by an efficient implementation (section 6.3).

### *6.2.1 Local Maximal Frequent Itemset Propagation in FastLMFI*

In FastLMFI approach we propagate a local index list (**LIND**) $LIND_{p+1}$ for each tail frequent itemset $FI_{p+1}$, which contains the indexes (positions) of those local maximal frequent itemsets in *list (MFI)*, in which $FI_{p+1}$ is appeared as a prefix. For example in Figure 10, node *A* contains the indexes of those local maximal frequent itemsets where *A* appeared as a prefix. Child $LIND_{p+1}$ of node *P* can be constructed by traversing indexes of parent $LIND_p$ and placing them into child $LIND_{p+1}$, which can be done in one step. Line 1 to 2 in Figure 11 shows the creation of $LIND_{p+1} = LIND_{p\ È\ \{i\}}$ in one step, where line 3 in Figure 11 at same time traverse indexes of parent $LIND_{p\ È\ \{i\}}$ and create child $LIND_{p+1} = LIND_{pÈ\ \{i\}}$ indexes.

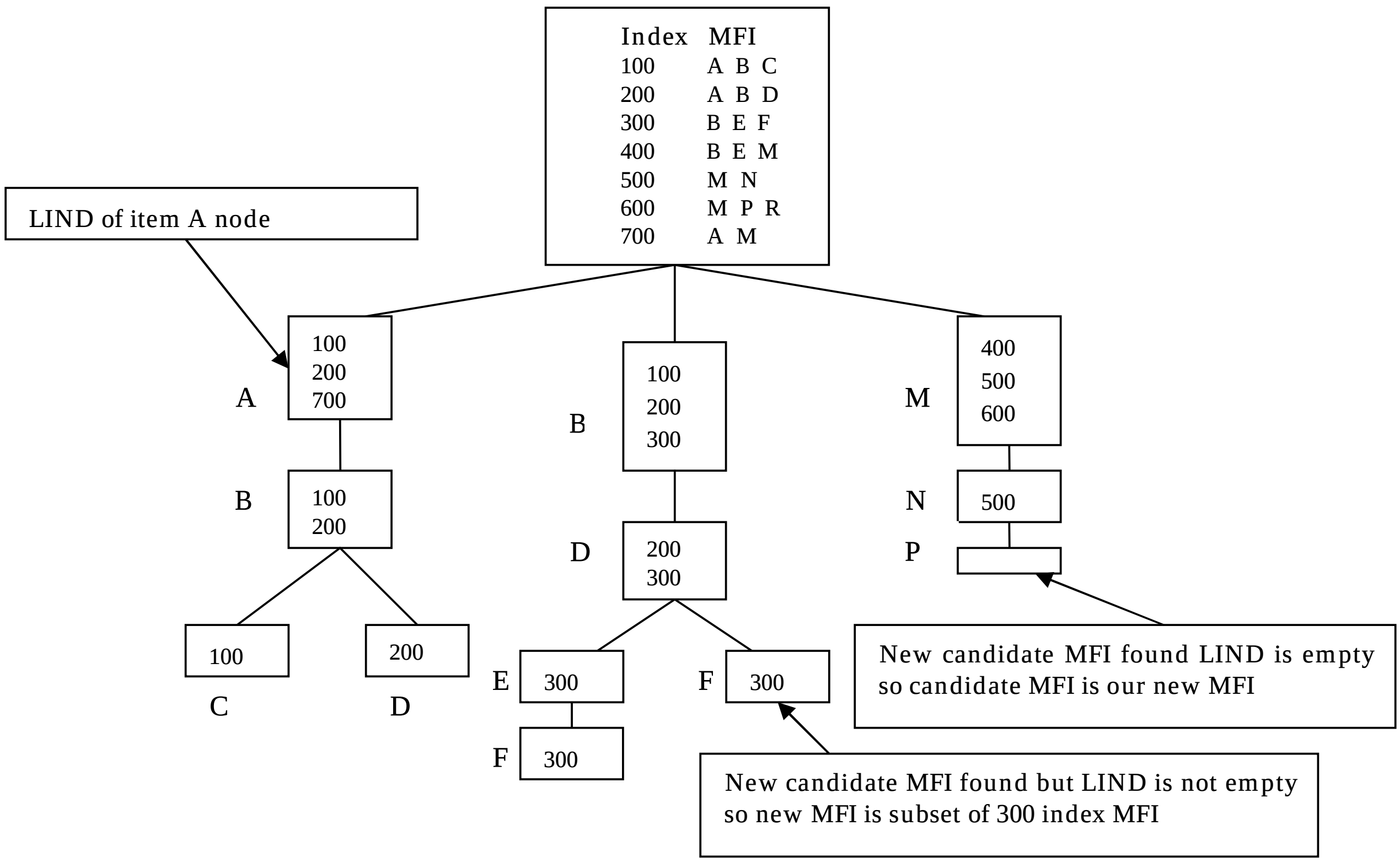

Figure 10: FastLMFI local maximal frequent itemsets (LMFI) propagation and itemset maxmality checking example.

```
Procedure PropagateLIND (LIND_p , P)

for each tail item of node P i ε tail (P)
        for each index of LIND_p
                LIND_p+1 = LIND_p È {i}
                PropagateLIND(LIND_p+1 , P È {i} )
IncrementSubtreeIndexes(LIND_p ,   LIND_p+1 )
```

Figure 11: Pseudo code of Local Maximal Itemset Propagation using FastLMFI approach.

```
Procedure IncrementSubtreeIndexes (LIND_p, LIND_p+1)

for each ℓ index of LIND_p+1 not in LIND_p
        LIND_p = LIND_p +ℓ
```

Figure 12: Pseudo code of incrementing parent local maximal itemset indexes.

**Lemma 1:** Let *P* be the node of search space and let $LIND_p$ contains its local maximal frequent itemsets. Its tail items $LIND_{p+1}$ can be constructed from local maximal frequent itemsets indexes of *P*.

**Proof:** We know that all tail items are *tail* $\{i\} \subseteq P$, and $LIND_p$ contains all those maximal frequent itemsets indexes, where *P* is appear as a prefix. So tail item $LIND_{p+1}$ can be constructed directly from indexes of $LIND_p$, because $LIND_{p+1} \subseteq LIND_p$.

Note that $LIND_p$ of itemset *P* contains exactly same number of local maximal frequent itemsets as progressive focusing $LMFI_p$. Only difference between the two techniques is that, our approach propagate an index list $LIND_{p+1}$ to child nodes, where progressive focusing pushes and places them in top or bottom of *list (MFI)*.

### *6.2.2 Incrementing Parent Local Indexes*

Note that node $LIND_p$ contains exactly those indexes of maximal frequent itemsets which are known to the parent of $LIND_p$. In other words $LIND_p$ does not contain those maximal frequent itemsets indexes which are mined later or found in subtree of *P*. To update those indexes found in subtree of P, we must add all new indexes of $LIND_{p+1}$ into $LIND_p$. Procedure *IncrementSubtreeIndexes (parent LIND, child LIND)* in Figure 12 shows the steps of incrementing parent indexes from its child node.

### *6.2.3 Itemset Maximality Checking*

If any node *P* finds a candidate maximal itemset, and if it contains an empty $LIND_p$, then the candidate maximal itemset will be our new mined maximal frequent itemset, otherwise it is subset of any $LIND_p$ itemset.

Figure 10 shows the process of propagation of $LIND_{p+1}$ and itemset maximality checking. Note that the root node contains all the known maximal frequent itemsets, which propagate $LIND_{p+1}$ to its child nodes.

***Example 1:*** Let us take an example of propagation of **LIND** from itemset *{A}* to itemset *{ABC}* of Figure 10 example. First, root node propagate itemset *A's* local maximal pattern indexes *{100,200,700}* to its child node *{A}*, because itemset *{A}* appears as a prefix in all these known maximal patterns. In next recursion, node *A* propagates local maximal pattern indexes to its child nodes, after comparing against its local maximal pattern indexes.  Itemset *{AB}* is appears as a prefix in *{100,200}* of node *A's* local maximal pattern indexes. Where itemset *{ABC}* is appears as a prefix in *{100}* of node *AB's* local maximal pattern indexes.

## 6.3 Efficient Implementation of FastLMFI

### *6.3.1 Maximal Frequent Itemset Representation*

We choose to use a vertical bitmap for the mined maximal patterns representation. In a vertical bitmap, there is one bit for each maximal pattern. If item i appears in maximal pattern j, then the bit of j of the bitmap of item i is set to one; otherwise the bit is set to zero. Figure 13 shows the vertical bit-vector representation of maximal patterns.

Note that each index of $LIND_p$ points to some position in *P = {0 È 1 È 2È … n}* bitmap. *P* child $LIND_{p+1}$ can be constructed by taking *AND* of $LIND_p$ bitmap, with tail item *X* bitmap.

**bitmap** ($LIND_{p+1}$) = **bitmap** ($LIND_p$) AND **bitmap**(*X*)

There are two ways of representing maximal patterns for each index of $LIND_p$. First, way is that each index of $LIND_p$ points to exactly one maximal pattern. Second, way can be each index of $LIND_p$ points to 32 maximal patterns of whole 32-bit integer range. The second approach was used for fast frequency counting in [8] and they show that it is better than single bit approach with a factor of 1/32. We also observed through experiments that second approach is more efficient than first approach for local maximal patterns propagation. Figure 14 compares the 32-maximal patterns per index with single maximal pattern per index, on retail dataset with different support thresholds.

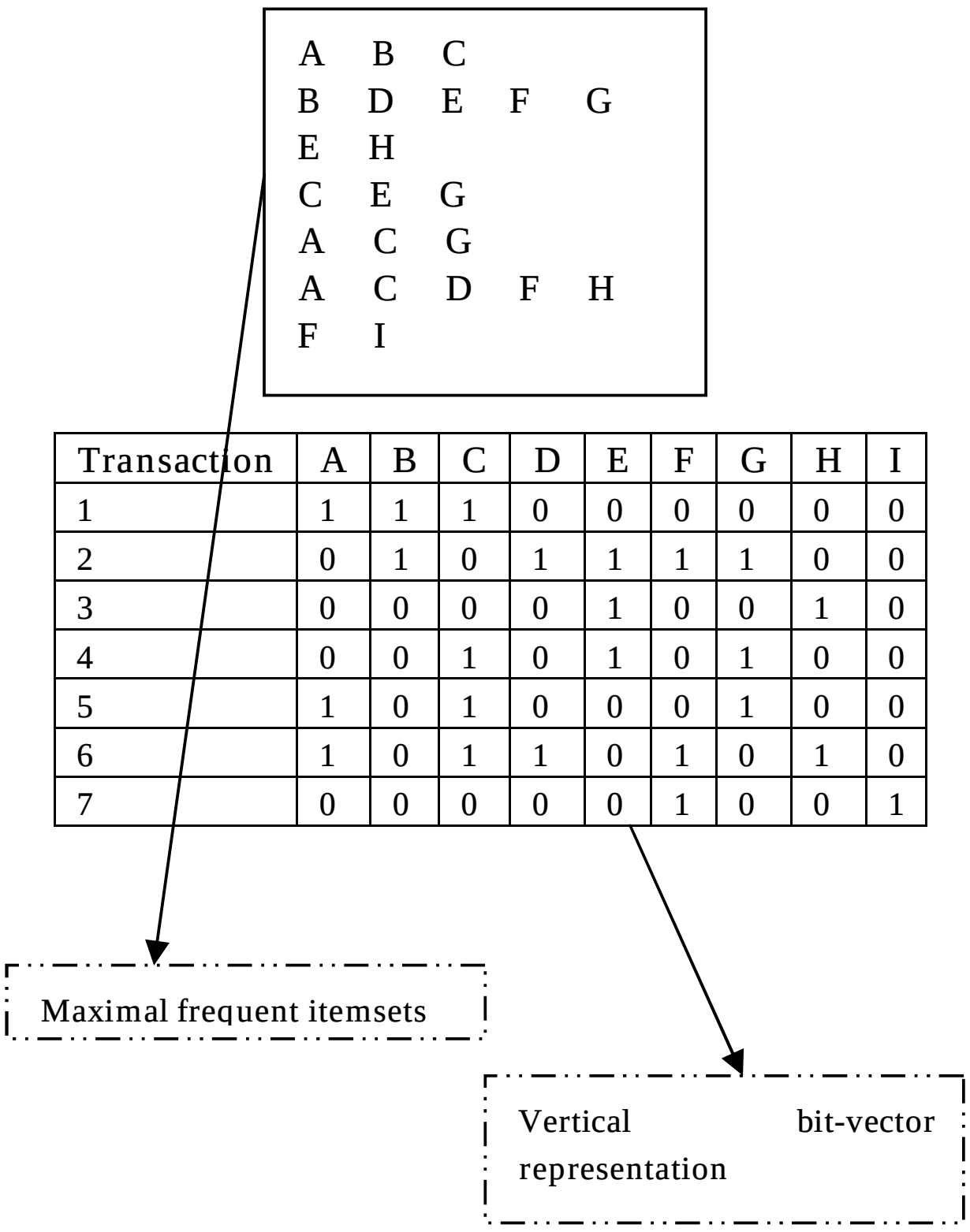

Figure 13: A sample maximal frequent itemsets representation using bit-vector approach.

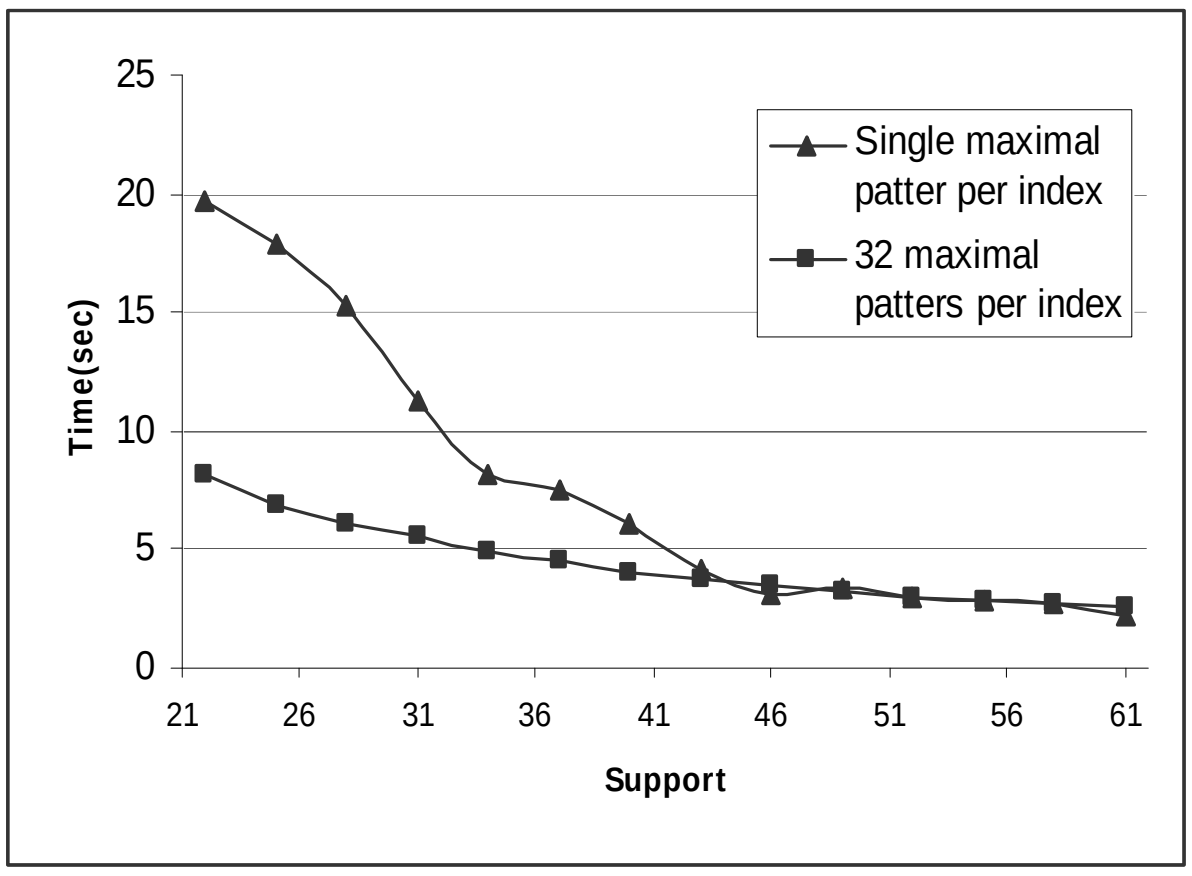

Figure 14: LIND Indexing with single maximal itemset versus Indexing with 32 maximal itemset.

### *6.3.2 Memory Optimization*

As explained earlier each recursion of **MFI** algorithm constructs and propagates $LIND_{p+1}$ to its child nodes. One way of construction of child $LIND_{p+1}$ is to declare a new memory and then propagate to child nodes. Obviously this technique is not space efficient. A better approach is as follows. We know that with Depth First Search (DFS) a single branch is explored at any time. Before starting the algorithm we create a large memory (equal to all known maximal patterns) for each level, which is equal to the maximal branch length. Next time each level of **DFS** tree can share this memory, and does not need to create any extra memory at each recursion level.

## 6.4 Ramp-max: Efficiently MFI Mining

The search strategy of *Ramp-max* for generating candidate itemset and counting frequency of these itemset is same as mining all frequent itemset. For removing non-maximal itemset we used three search space pruning techniques PEP, FHUT, and FHUTMFI. These techniques are described in [8], while we used **FastLMFI** for maximal itemset super set checking instead of progressive focusing. The pseudo code of *Ramp*-max is described in Figure 15.

```
Ramp-max (Node n, IsHUT)

(1)   HUT = P.head ∪ P.tail
(2)   If HUT is in list (MFI)
(3)       stop and return

(4)   for each item X in n.tail
(5)       for each region index ℓ in PBR<n>
(6)           AND-result = bit-vector[ℓ] ∧ head-
              bit-vector of n [ℓ]
(7)           Support[X] = support[X] + number of
              ones(AND-result)

(8)  Use PEP to trim the tail. Remove infrequent items
     from n.tail and reorder them by increasing
     support

(9)  for each item X in n.tail
(10)      m.head = n.head È X
(11)      m.tail = n.tail – X
(12)      for each region index ℓ in PBR<n>
(13)          AND-result = bit-vector[ℓ] ∧ head-bit-
              vector[ℓ]
(14)          If AND-result > 0
(15)              Insert ℓ in PBR<m>
(16)              head bit-vector of m [ℓ] = AND-result
(17)      Ramp-max (m)
(18)  If (IsHUT and all extensions are frequent)
(19)      Stop search and go back up subtree
(20)  If (P is a leaf and P.head is not in list (MFI))
(21)      Add C.head to list (MFI)
```

Figure 15: Ramp-max: Pseudo code for mining maximal frequent itemset using PBR projection technique.

```
Ramp-closed (Node n)

(1)  for each item X in n.tail
(2)    for each region index ℓ in PBR<n>
(3)        AND-result = bit-vector[ℓ] ∧ head- bit-vector
           of n [ℓ]
(4)        support[X] = support[X] + number of
           ones(AND-result)

(5)  Remove infrequent items from n.tail and reorder
     them by increasing support

(6)  for each item X in n.tail
(7)      m.head = n.head È X
(8)      m.tail = n.tail – X
(9)      for each region index ℓ in PBR<n>
             AND-result = bit-vector[ℓ] ∧ head-bit-
             vector[ℓ]
(10)           If AND-result > 0
(11)               Insert ℓ in PBR<m>
(12)               head bit-vector of m [ℓ] = AND- result
(13)     Ramp-closed (m)

(14)  If all of m superset itemset have support less than
      support (m)
(15)      Add m in list (CFI)
```

Figure 16: Ramp-closed: Pseudo code for mining closed frequent itemset using PBR projection technique.

## 7 MINING CLOSED FREQUENT ITEMSETS

As we presented earlier, an itemset is closed, if none of its superset is as frequent as itself. Our Ramp-closed implementation is almost same as Ramp-max. The only big difference is that, in Ramp-max an itemset is maximal if its node's **LIND** found empty, while in Ramp-closed we check the support of each known closed frequent itemset in **LIND**. If the support of all itemsets is less than *support (X)* then X is declare as closed itemset. The pseudo code of *Ramp*-closed is described in Figure 16.

# 8 EXPERIMENTAL RESULTS

In this section, we show the results of our computational experiments which we performed on different benchmark datasets. The implementations of Ramp-(all/max/closed) are coded in C language, and the experiments are done on Pentium4 3.2 GHz CPU with 512MB memory. The performance of Ramp-all, Ramp-max and Ramp-closed are compared with the current best algorithms which marked good score on FIMI03 and FIMI04 [11]. Due to the lack of space, we can not show the experimental results with all datasets. Therefore we classified the datasets into four different groups and select two dataset from each group. Figure 17 shows main features of our experimental datasets.

Our first group is composed of BMS-WebView1, BMS-WebView2 and Retail datasets. These datasets have many items but small number of transactions and are sparse. We choose BMS-WebView1 and BMS-WebView2 for performance comparison. Our second group is composed of BMS-POS and Kosarak datasets. These datasets have many items and also large number of transactions. If the minimum support is given to very small, these datasets generates huge number of frequent itemsets. We choose BMS-POS and Kosarak for performance comparison. Our third group is composed of Chess, Connect, Pumsb, Pumsb-star, accidents and Mushroom datasets. These datasets are very dense and almost 90% of time is spend on writing frequent itemsets to output file, if the minimum support is given to very small. We choose Mushroom and Chess for performance comparison. Our last group is composed of T10I4D100K and T40I10D100K. These datasets are also very sparse and have large number of items. We select both datasets for performance comparison.

## 8.1 Algorithms used for Performance Comparisons

For performance comparison we used the original implementations of AFOPT [19], Fpgrowth-(zhu) [14] and MAFIA [8] provided by their respective authors. These all implementations can be downloaded from http://fimi/implementations.html.

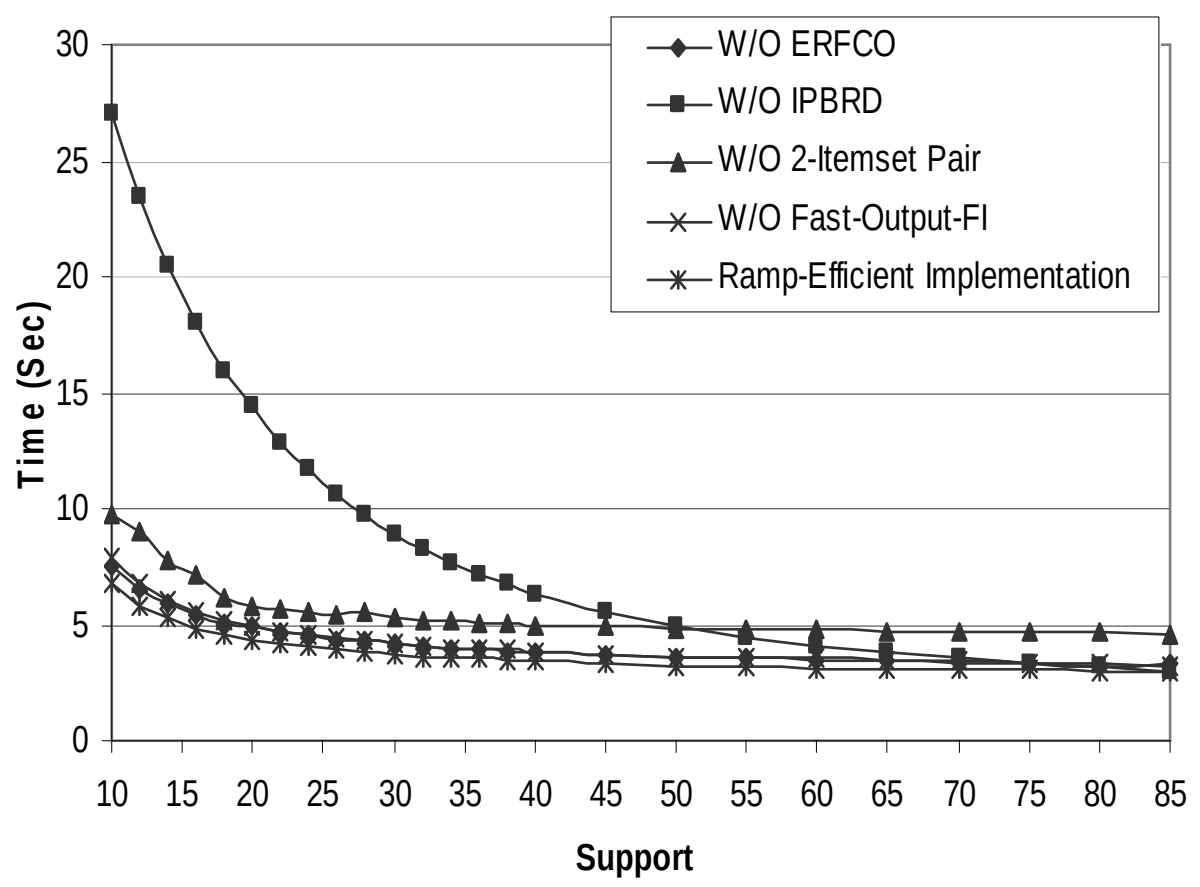

Figure 17: Performance analysis of ERFCO, IPBRD, 2-Itemset Pair and Fast-Output-FI on T10I4D100K dataset.

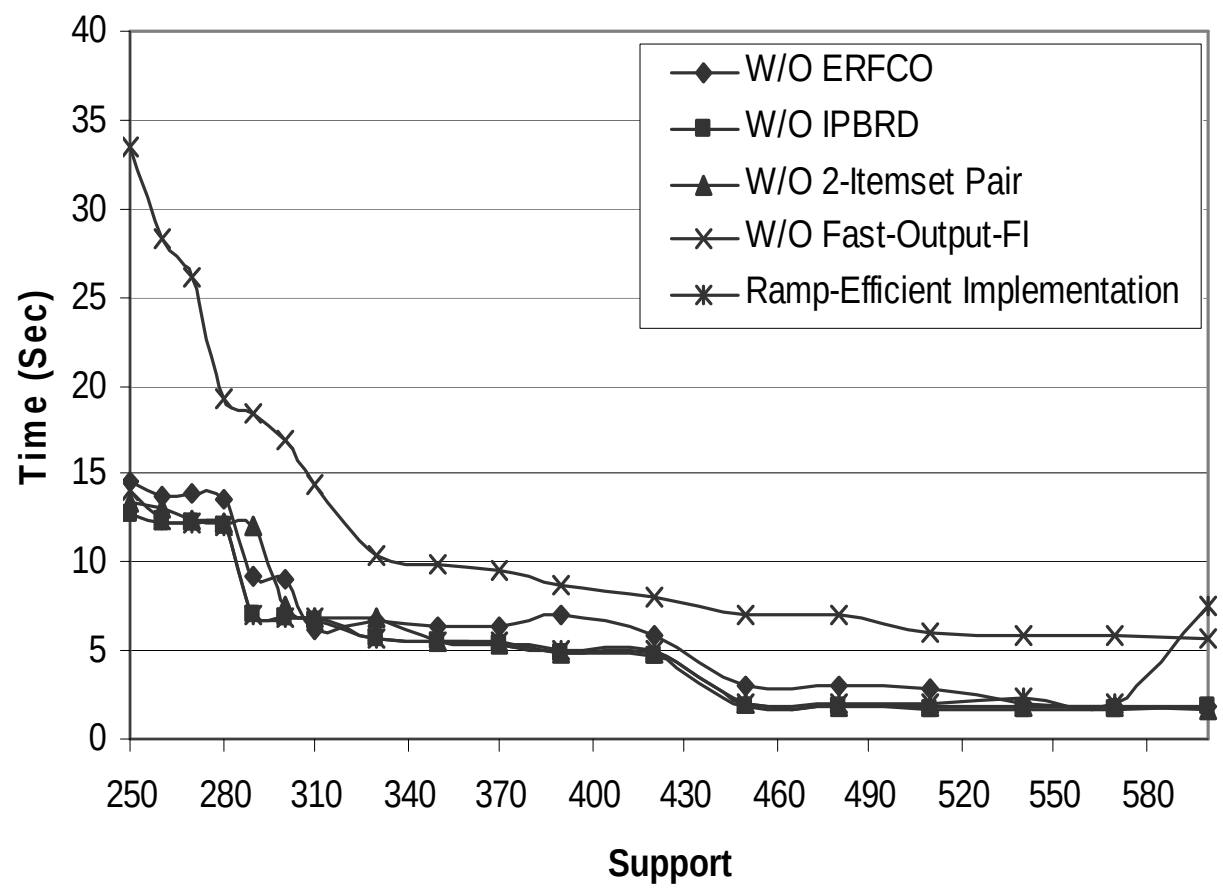

Figure 18: Performance analysis of ERFCO, IPBRD, 2-Itemset Pair and Fast-Output-FI on Mushroom dataset.

## 8.2 Performance Analysis of Ramp Components

In this section, we showed the performance analysis and effect of each component of Ramp algorithm on sparse and dense type datasets. The simulations are performed only on mining all frequent itemset problem, but we sure that each component of the Ramp, also performs exactly same on **MFI** and **FCI** mining problems. Our first experiment is on T10I4D100K, which is sparse dataset. Figure 17 shows the experimental results on T19I4D100K dataset. From the Figure 17, it is clear that **IPBRD** has the biggest effect on low level support thresholds as compare to other three components of Ramp-all implementation. On higher level threshold, we paid some extra cost of **2-Itemset pair** checking, but as the support threshhold decreases, **2-Itemset pair** checking improves the overall performance of algorithm. Our second experiment is on Mushroom, which is dense dataset. Figure 18 shows the experimental results on Mushroom dataset. As we have already explain, that on dense type datasets almost 90% of time in spend on writing output frequent itemsets to text file. Since in our Ramp-all implementation, we introduce a new frequent itemset writing technique **Fast-Output-FI**, so it has the biggest effect as compare to other three techniques. On dense datasets we noticed that **IPBRD** and **2-Itemset pair** does not improve the performance of algorithm, and without these components the Ramp-all sometimes gives good performance.

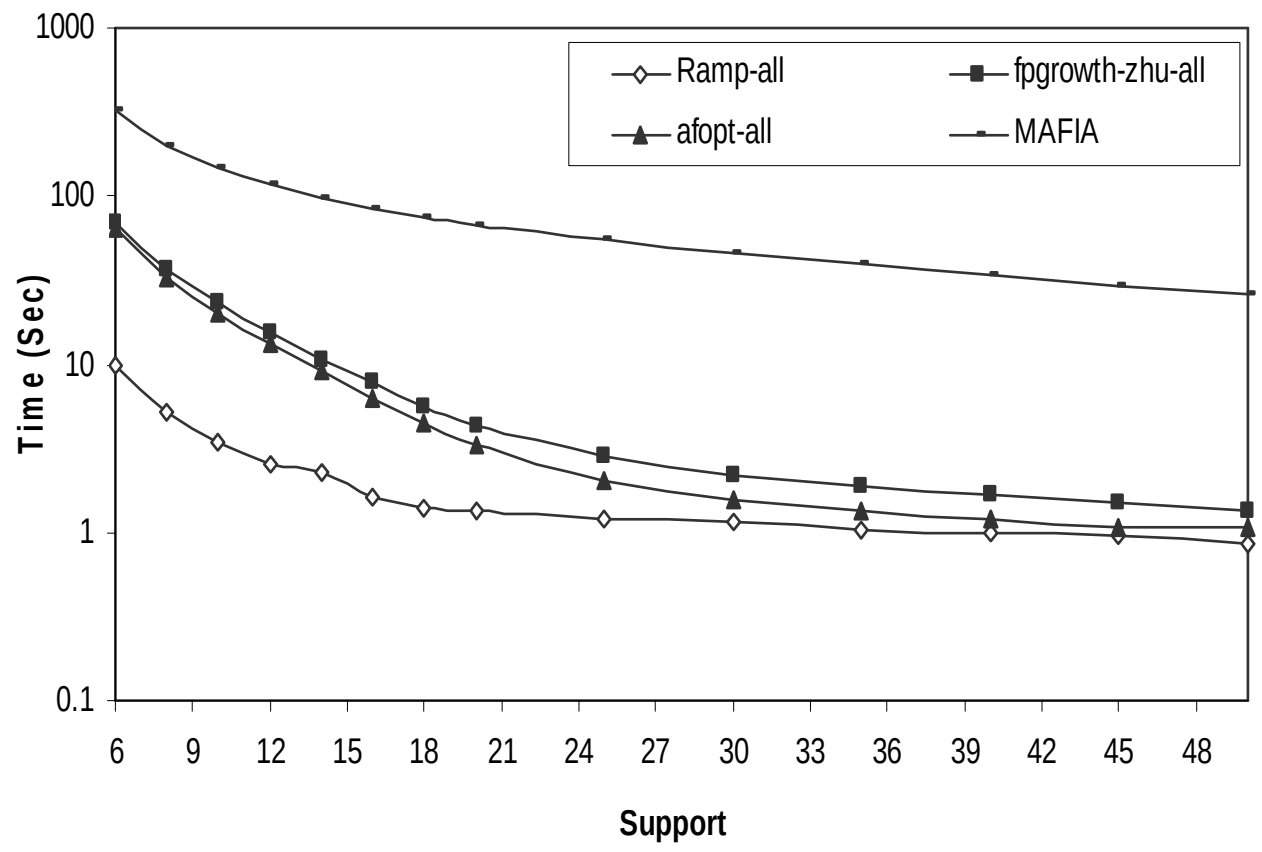

Figure 19: Performance results of Ramp-all on BMS-WebView2 dataset.

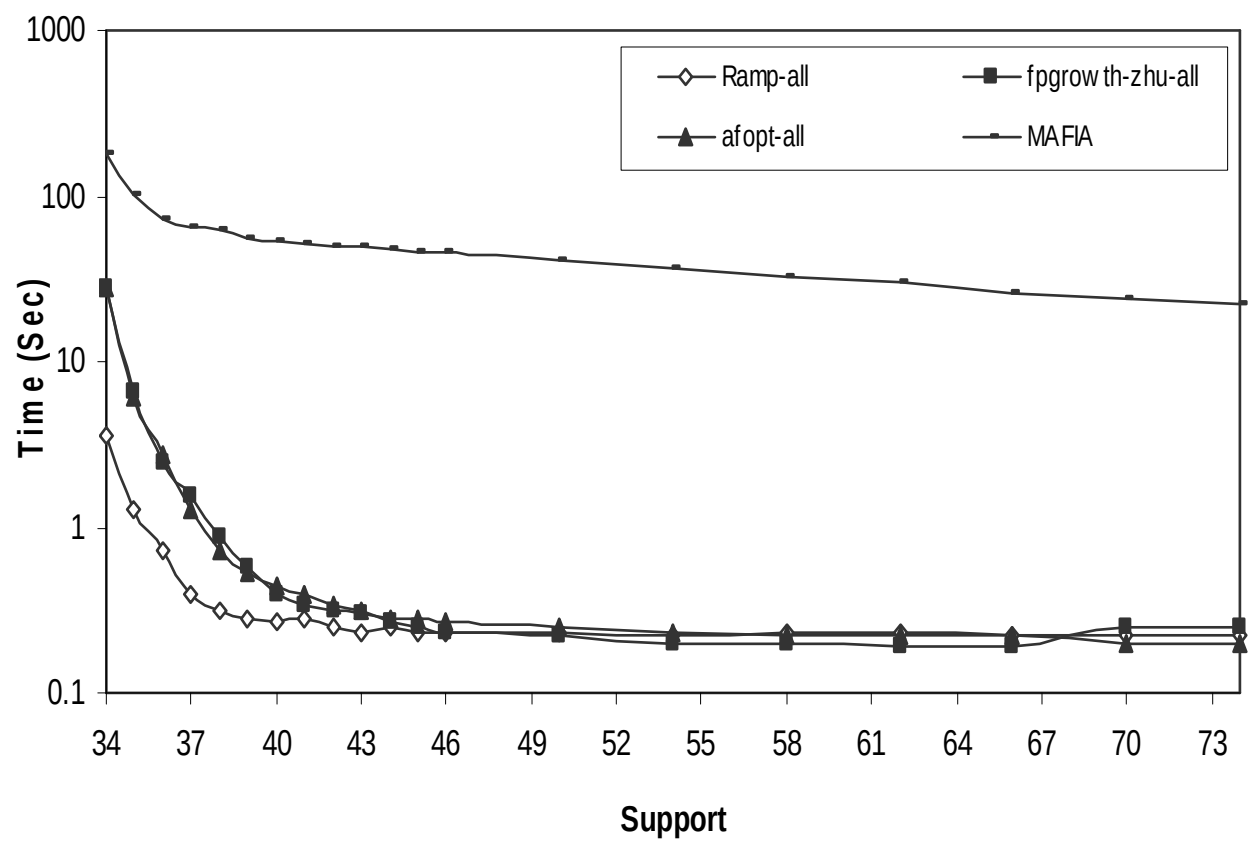

Figure 20: Performance results of Ramp-all on BMS-WebView1 dataset.

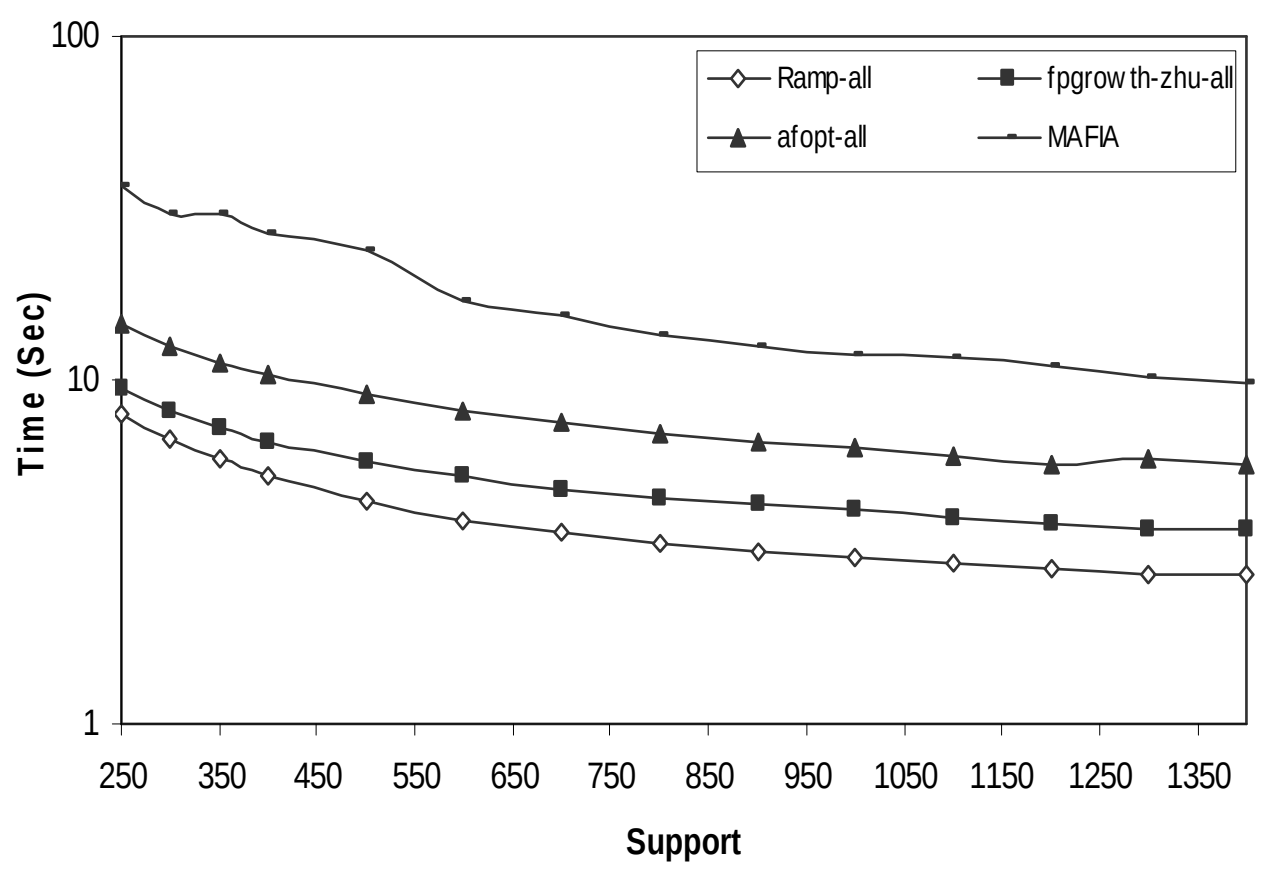

Figure 21: Performance results of Ramp-all on BMS-POS dataset.

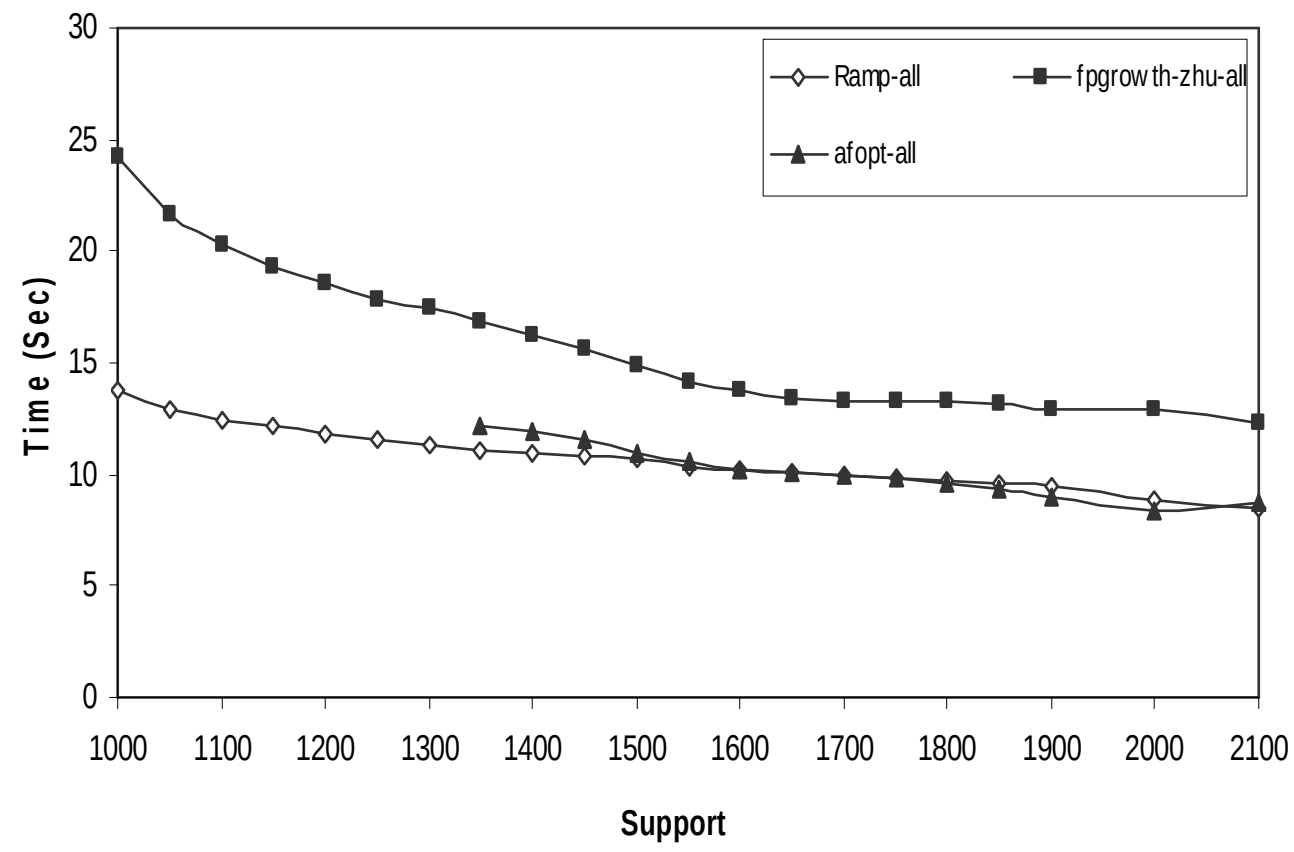

Figure 22: Performance results of Ramp-all on Kosarak dataset.

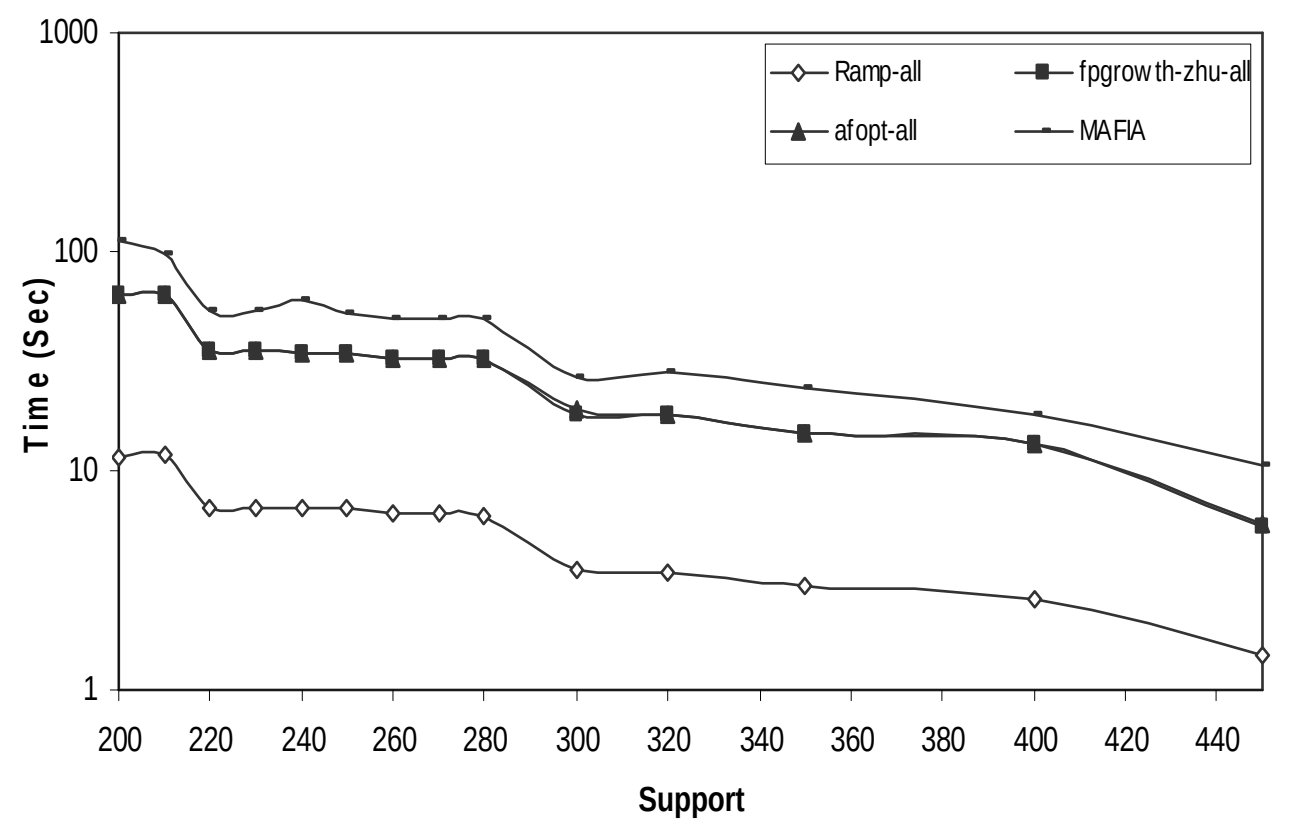

Figure 23: Performance results of Ramp-all on Mushroom dataset.

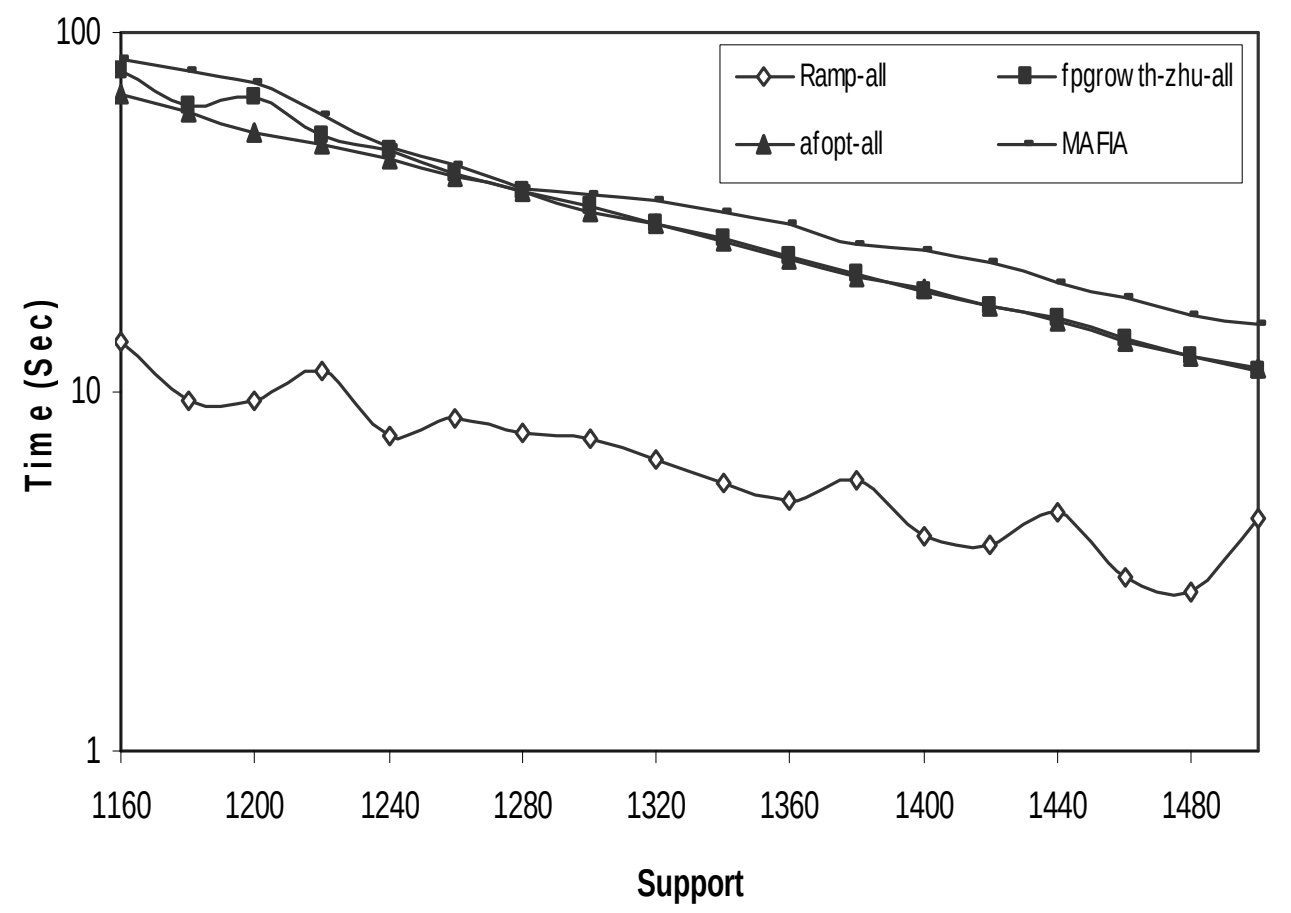

Figure 24: Performance results of Ramp-all on Chess dataset.

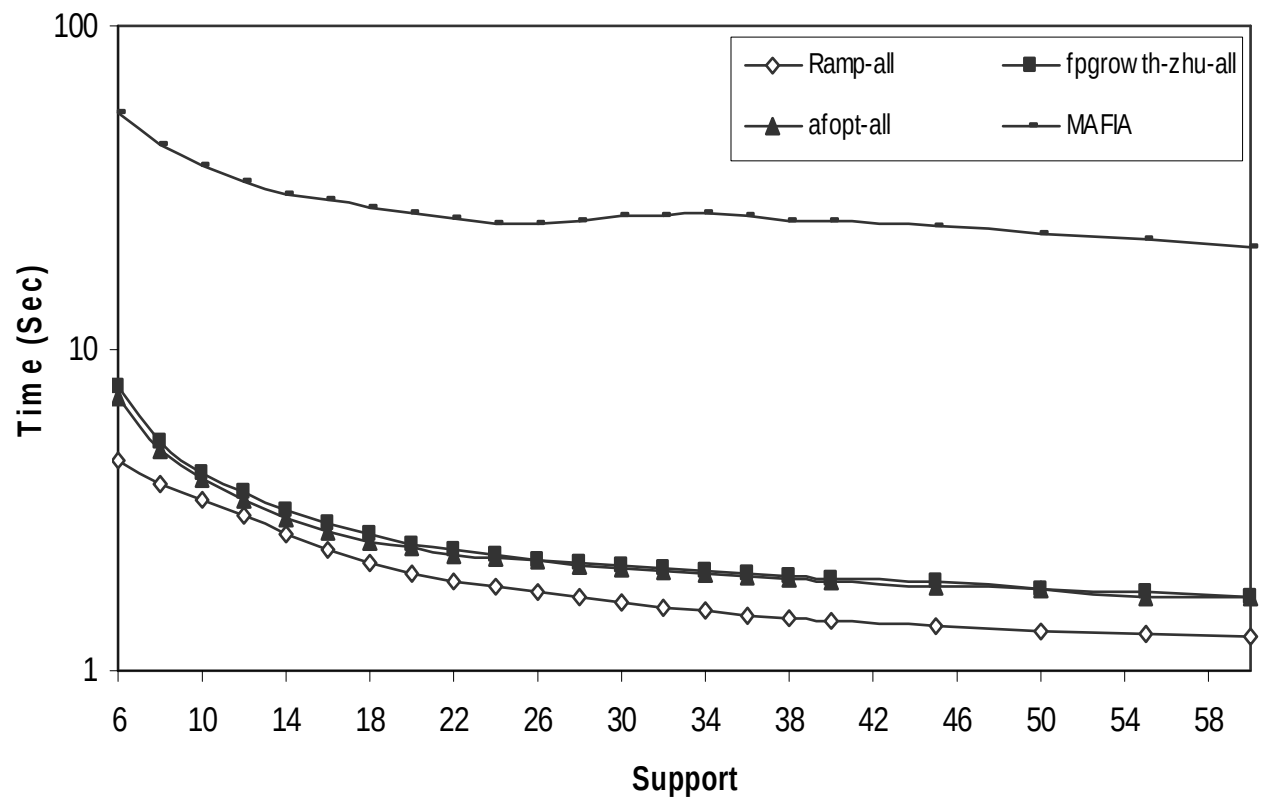

Figure 25: Performance results of Ramp-all on T10I4D100K dataset.

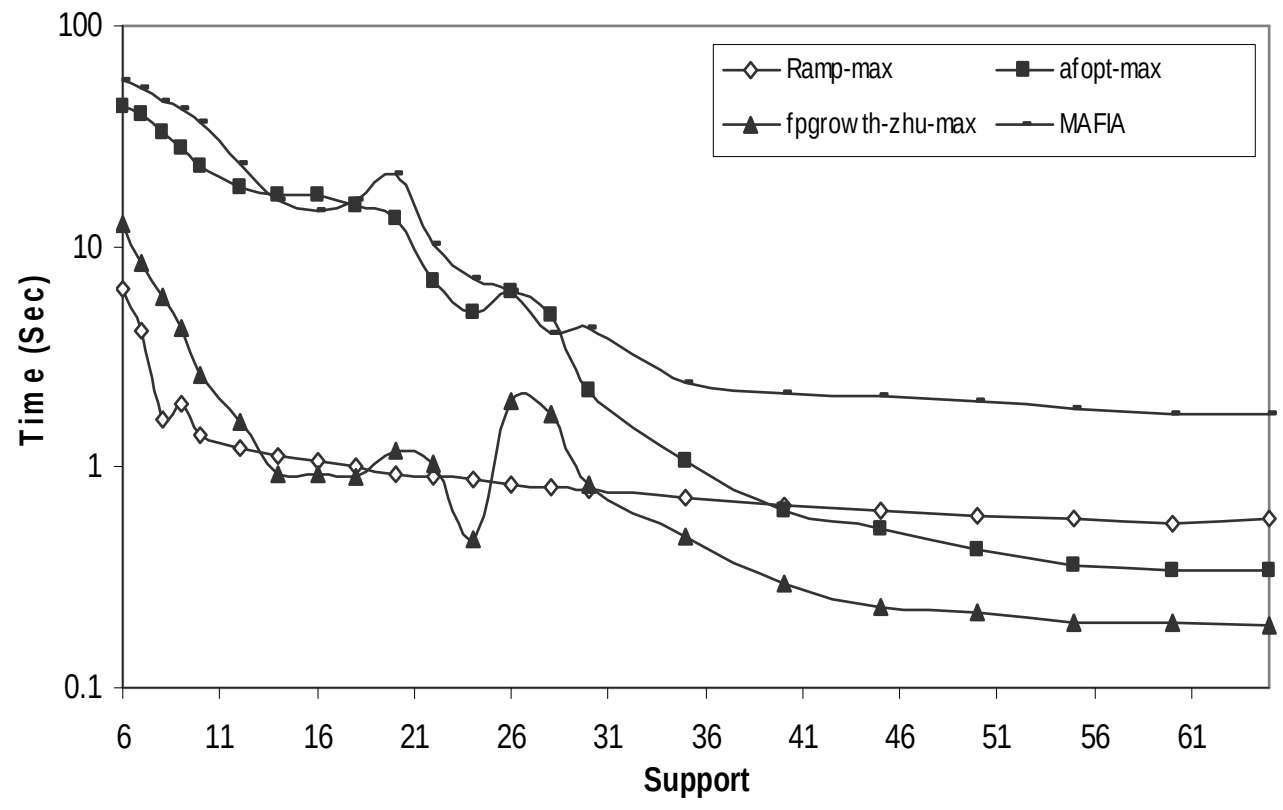

Figure 28: Performance results of Ramp-max on BMS-WebView1 dataset.

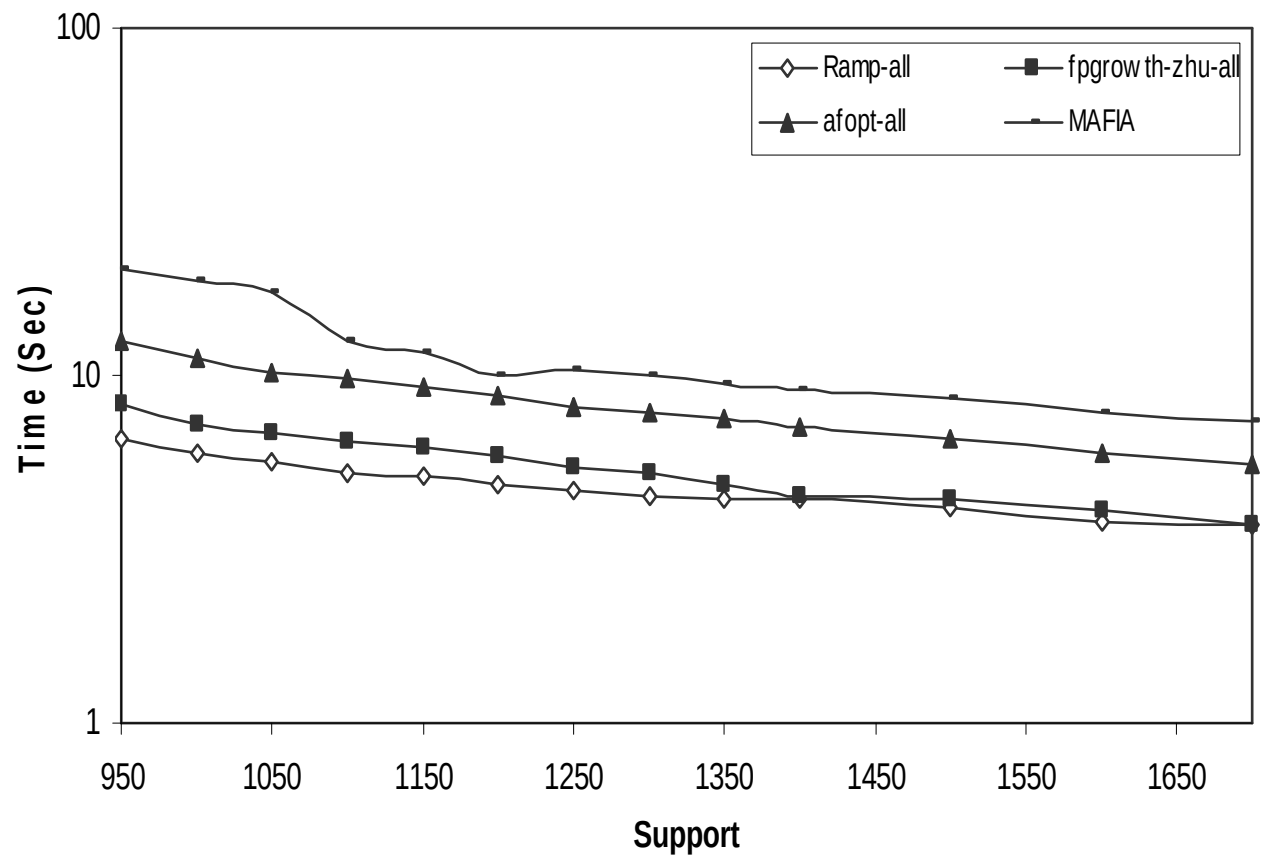

Figure 26: Performance results of Ramp-all on T40I410D100K dataset.

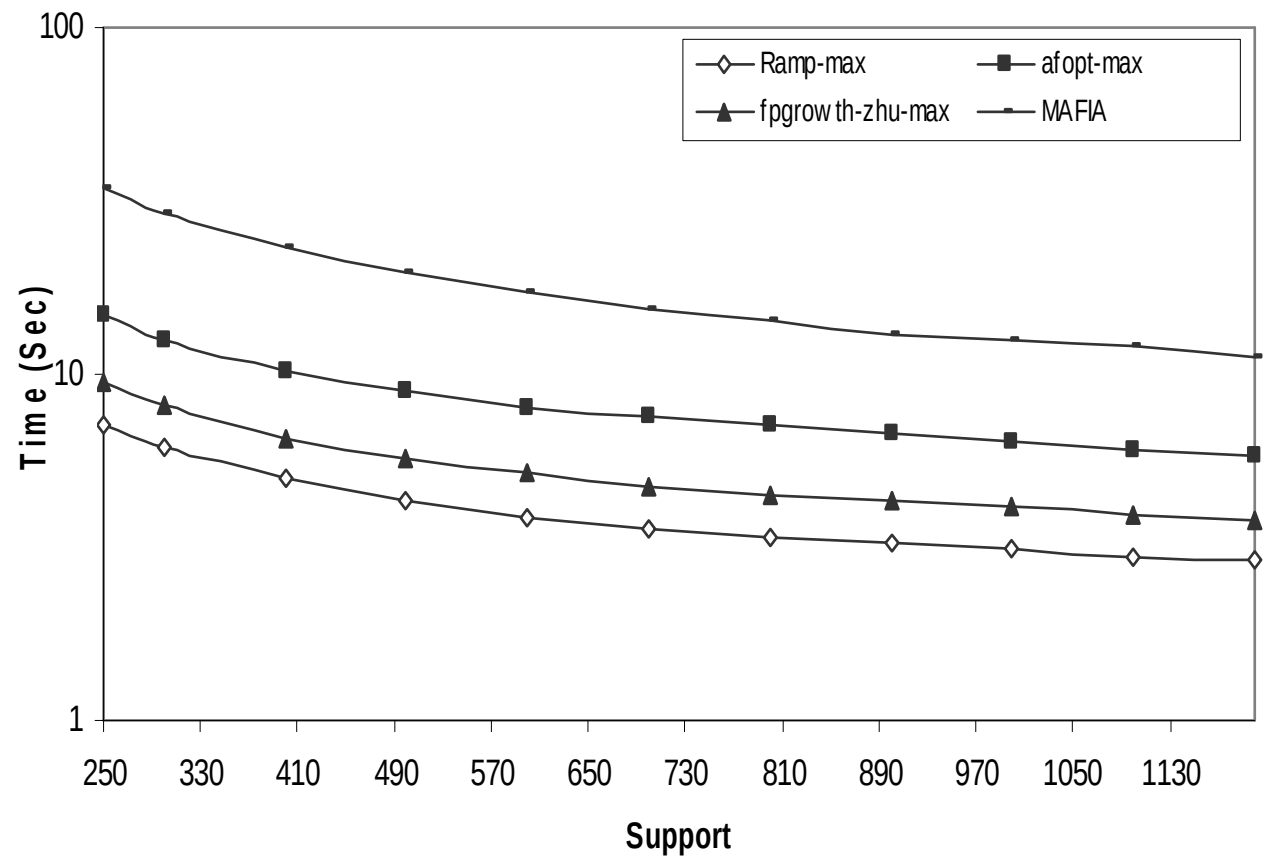

Figure 29: Performance results of Ramp-max on BMS-POS dataset.

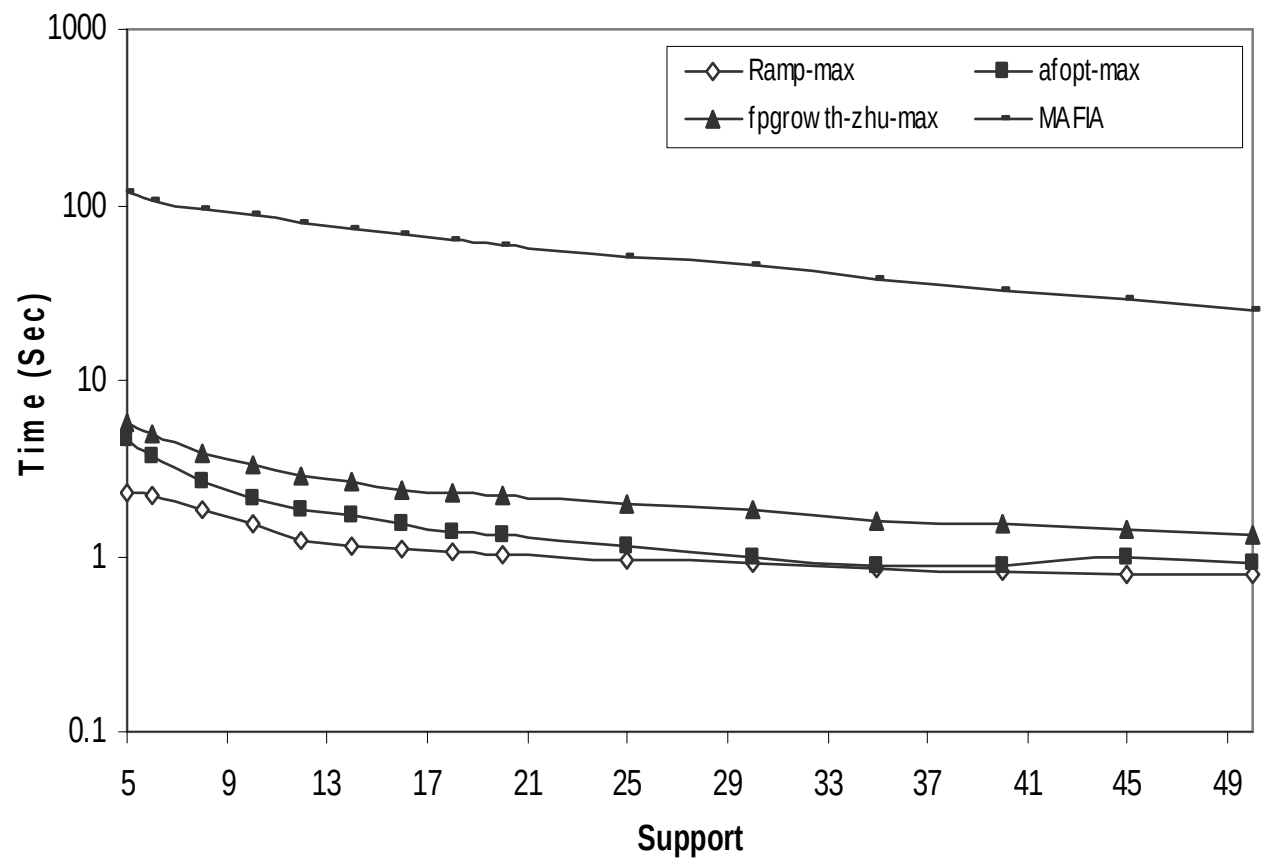

Figure 27: Performance results of Ramp-max on BMS-WebView2 dataset.

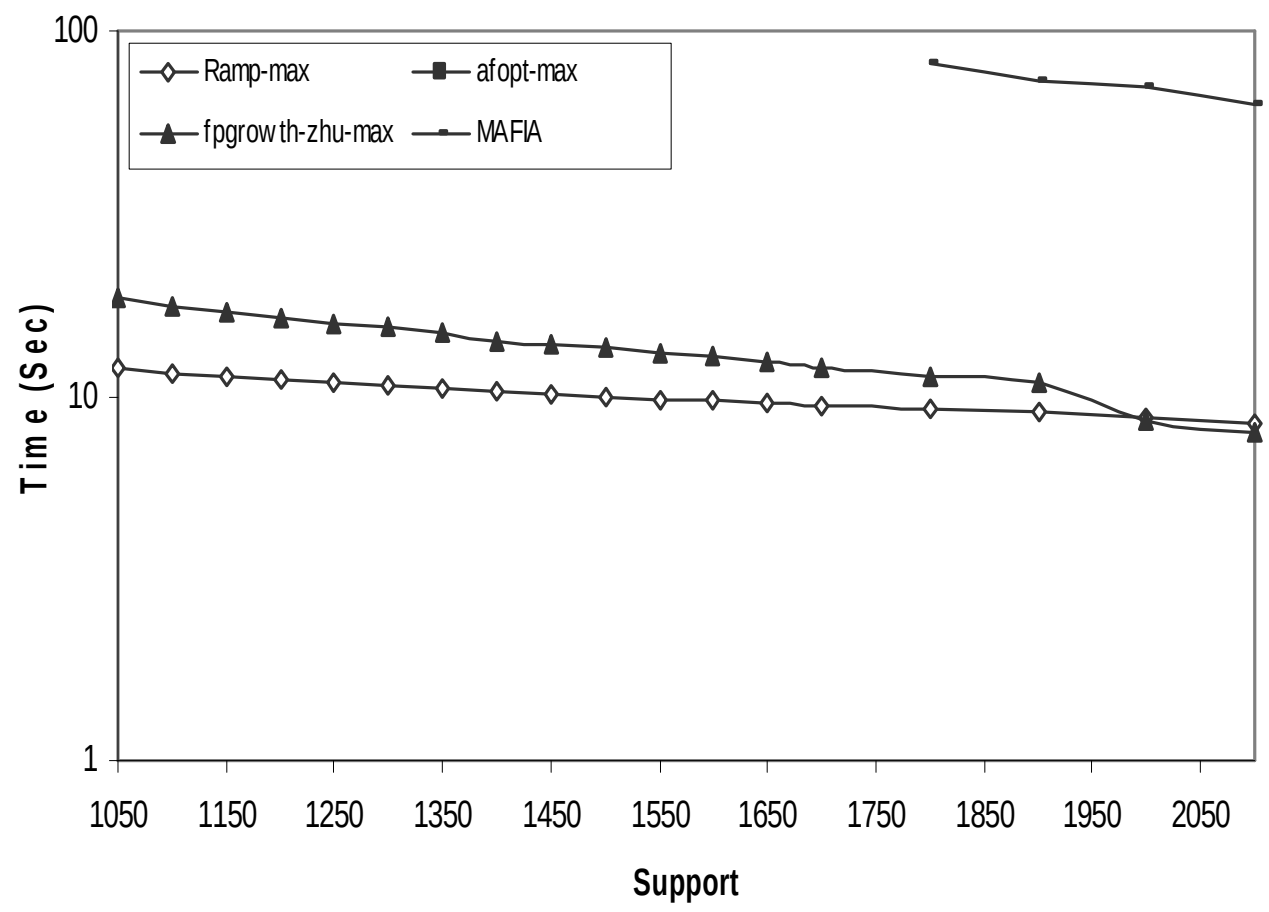

Figure 30: Performance results of Ramp-max on Kosarak dataset.

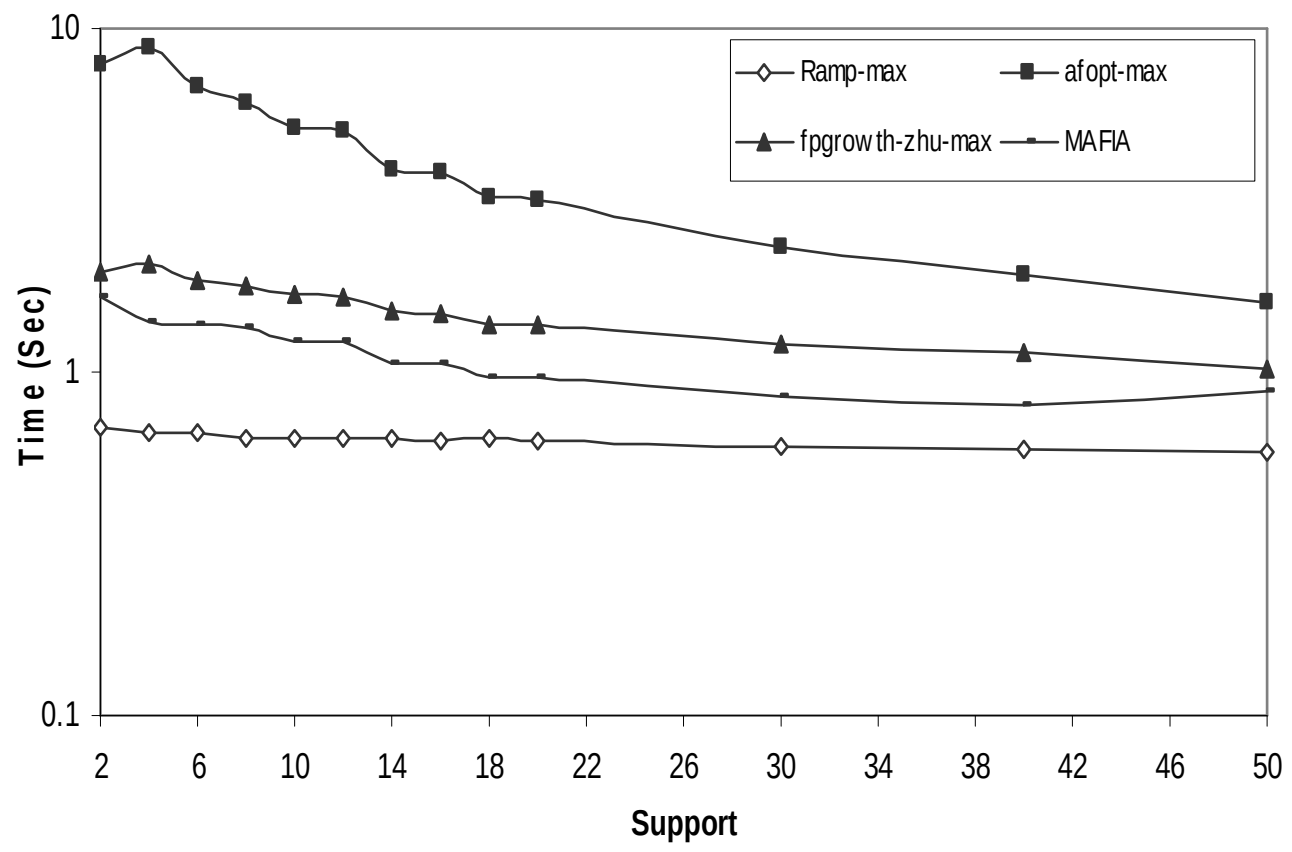

Figure 31: Performance results of Ramp-max on Mushroom dataset.

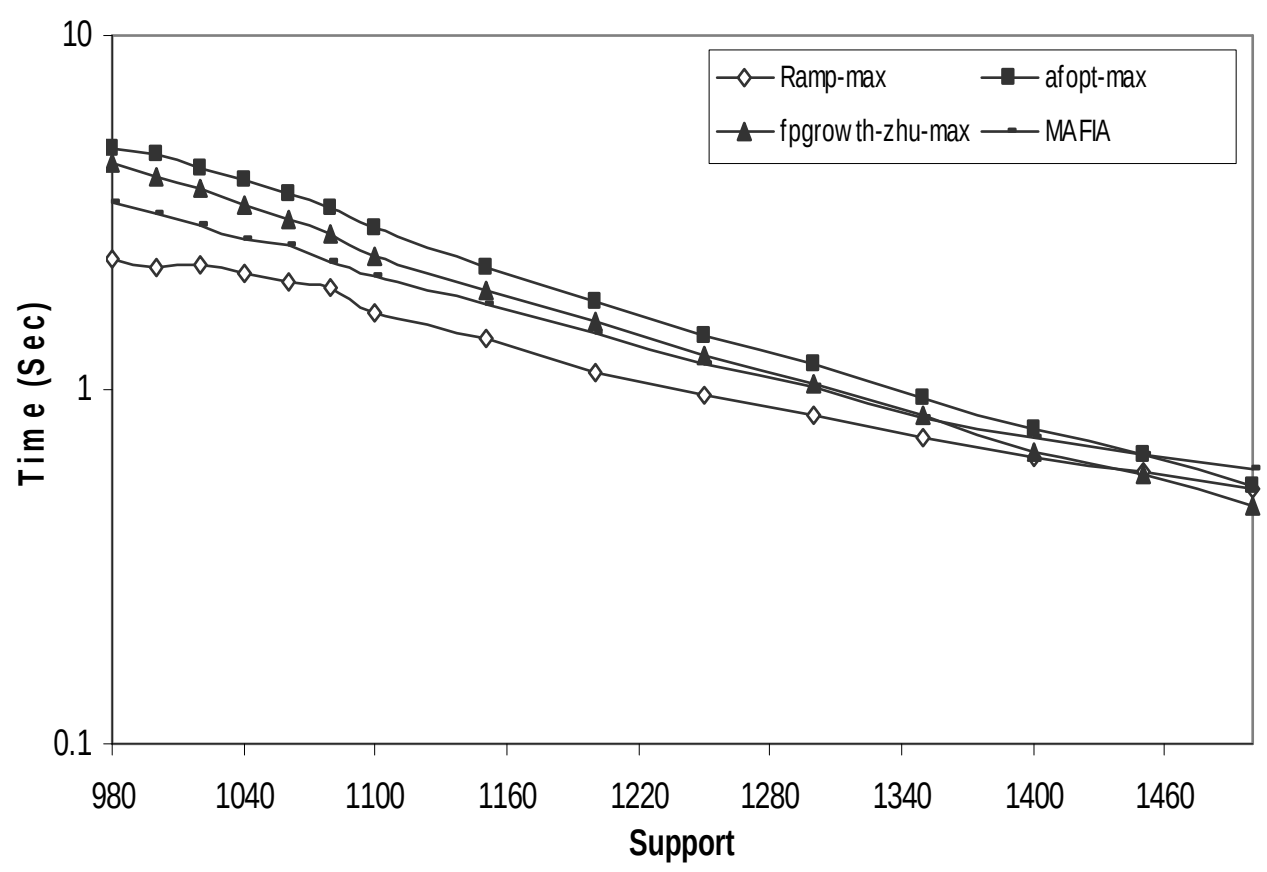

Figure 32: Performance results of Ramp-max on Chess dataset.

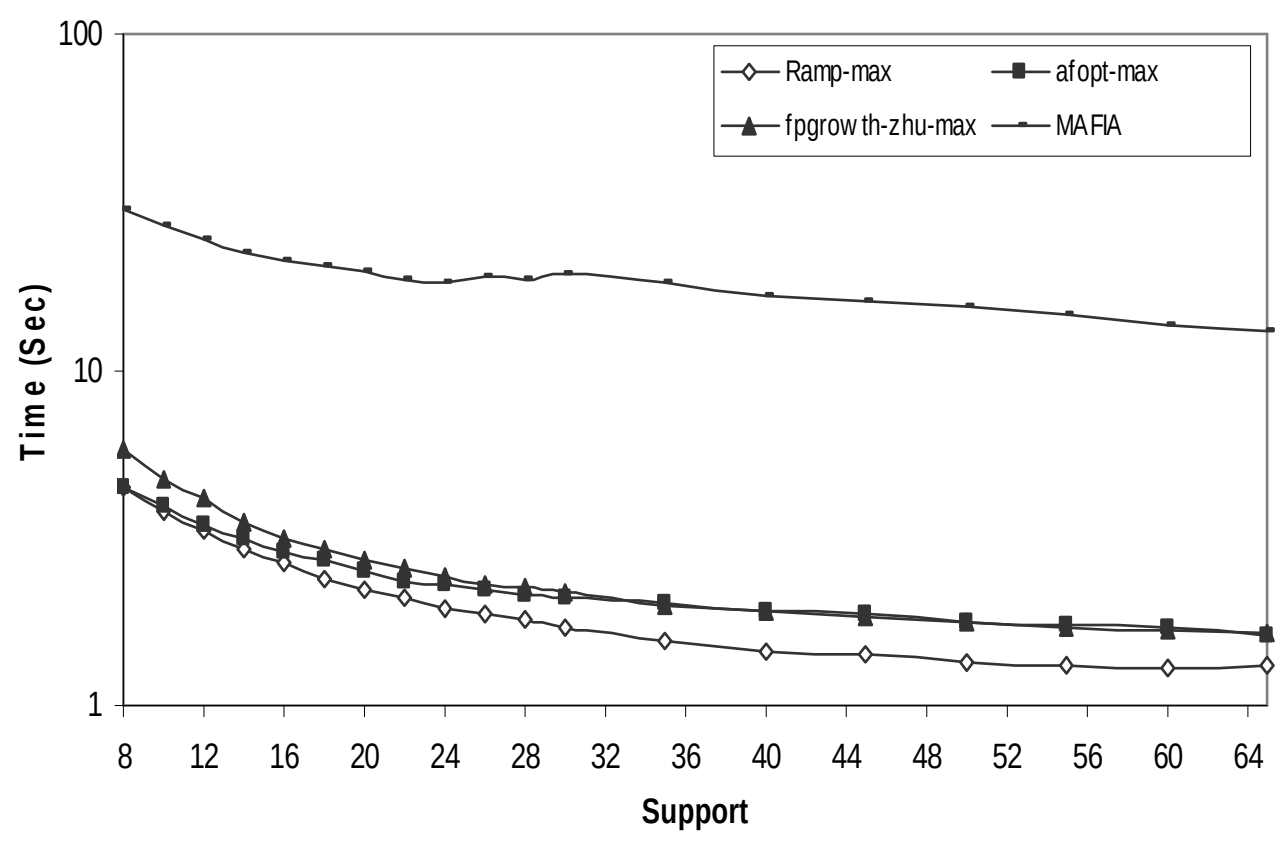

Figure 33: Performance results of Ramp-max on T10I4D100K dataset.

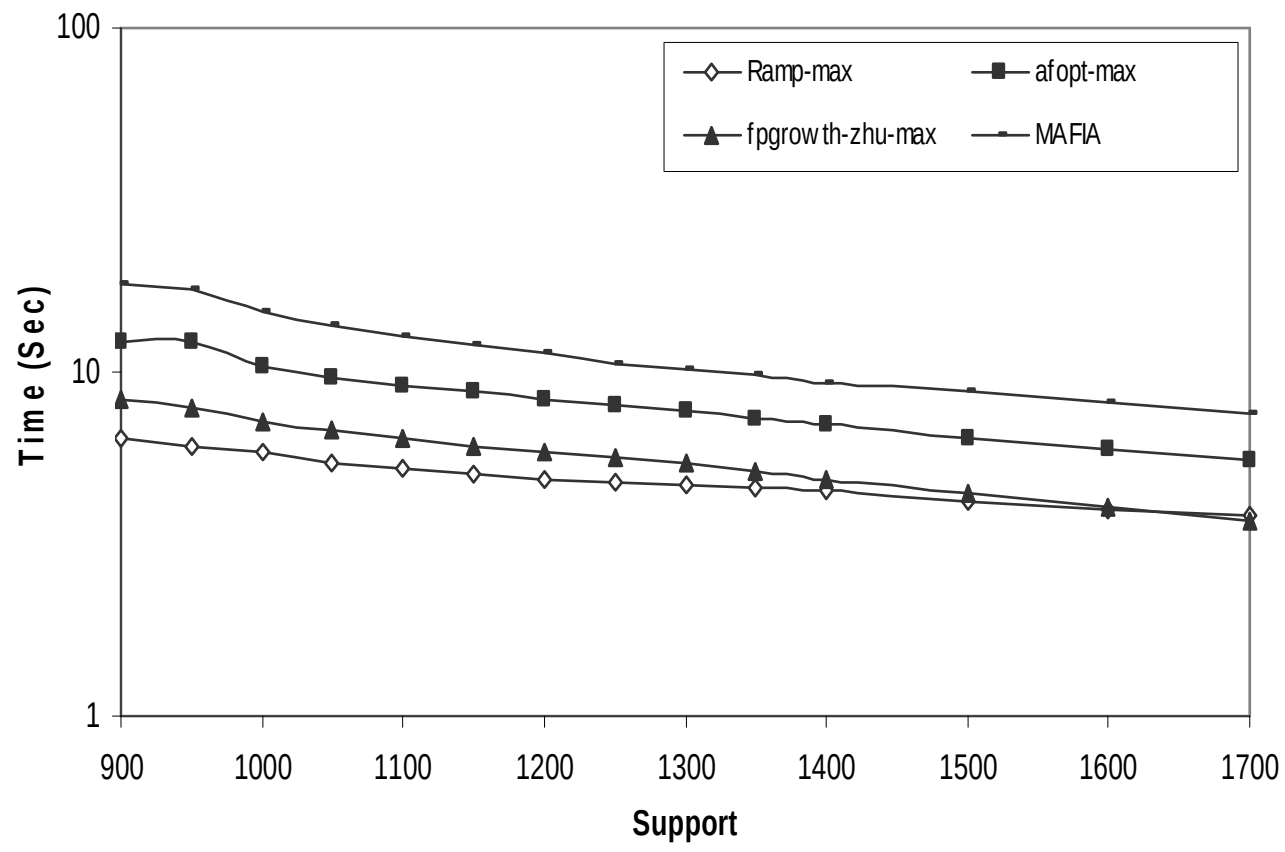

Figure 34: Performance results of Ramp-max on T40I410D100K dataset.

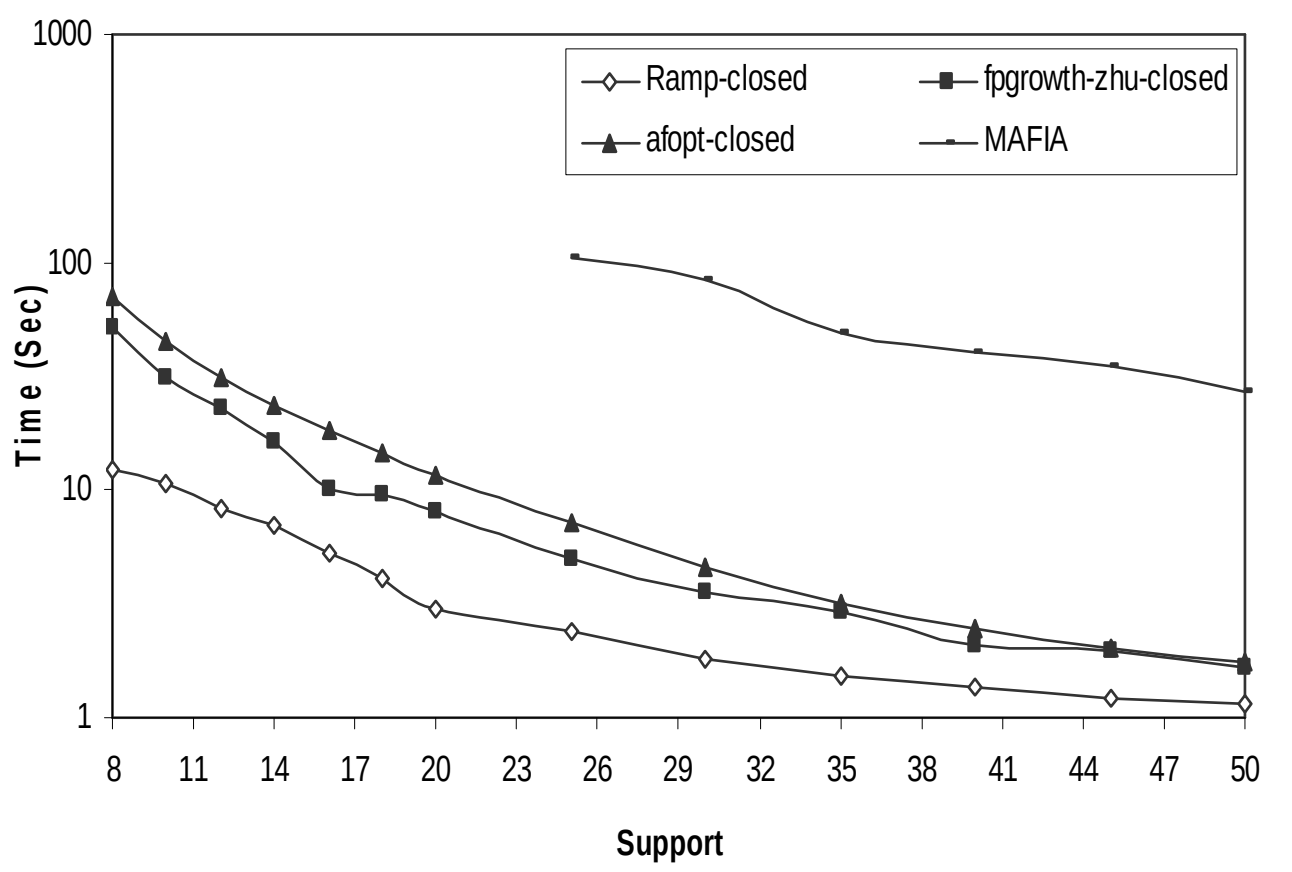

Figure 35: Performance results of Ramp-closed on BMS-WebView2 dataset.

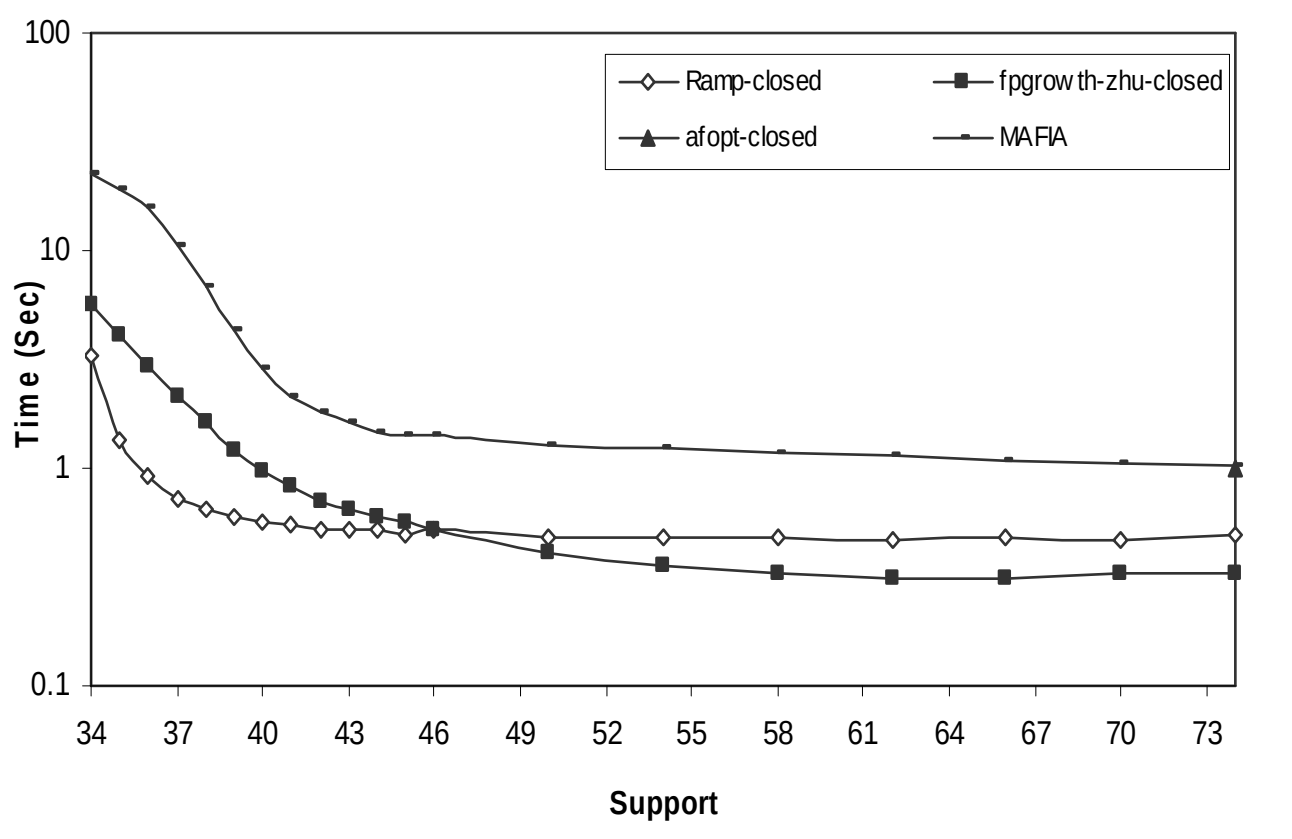

Figure 36: Performance results of Ramp-closed on BMS-WebView1 dataset.

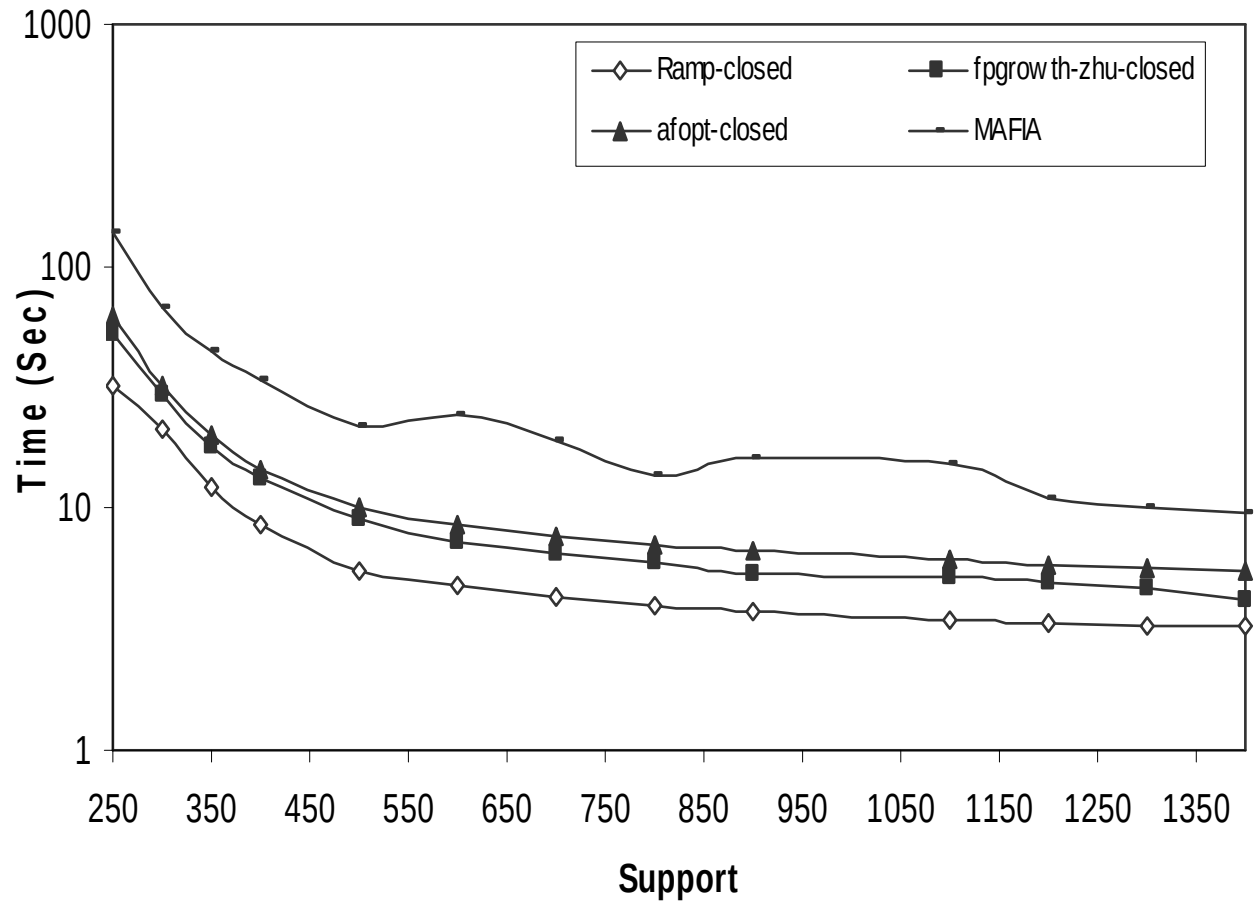

Figure 37: Performance results of Ramp-closed on BMS-POS dataset.

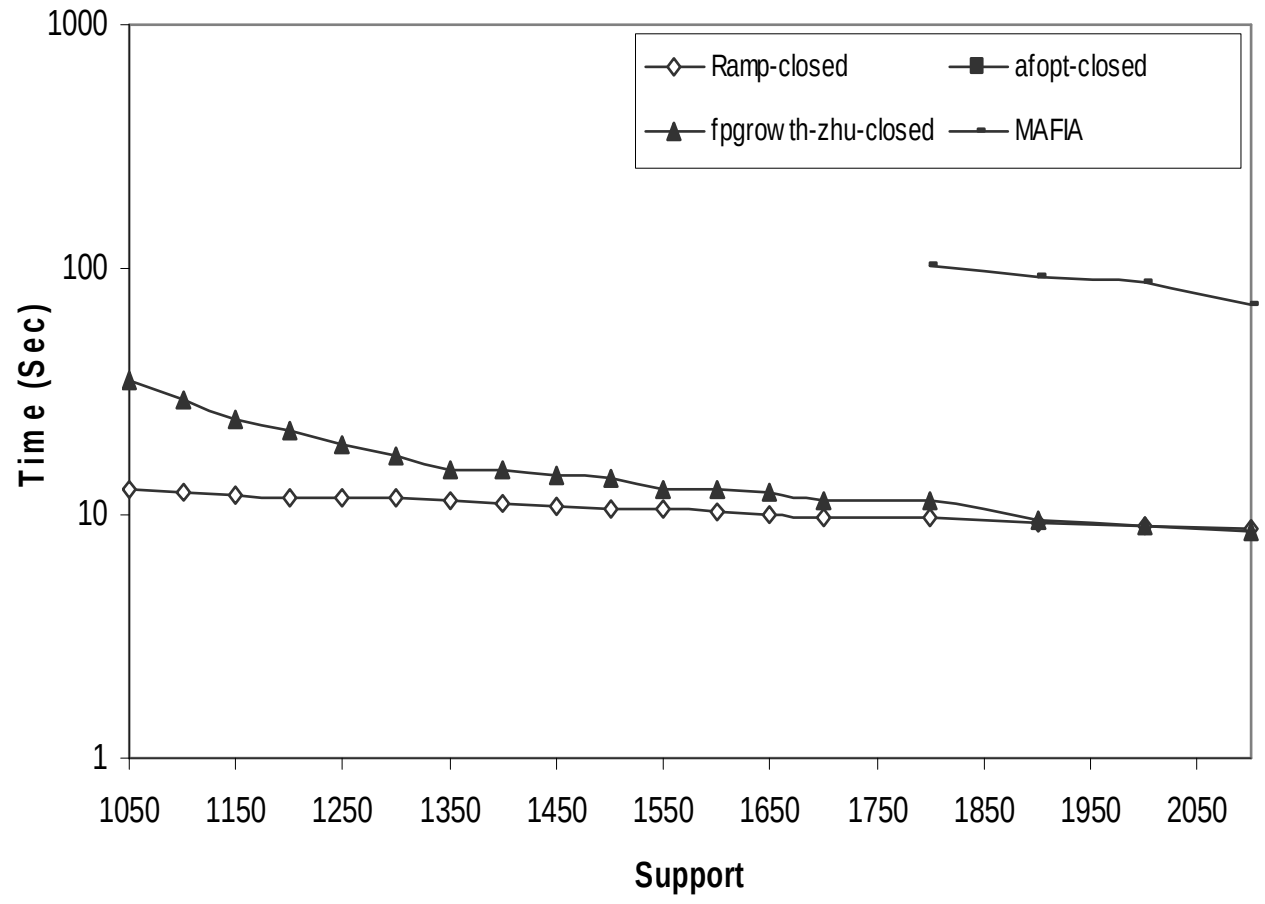

Figure 38: Performance results of Ramp-closed on Kosarak dataset.

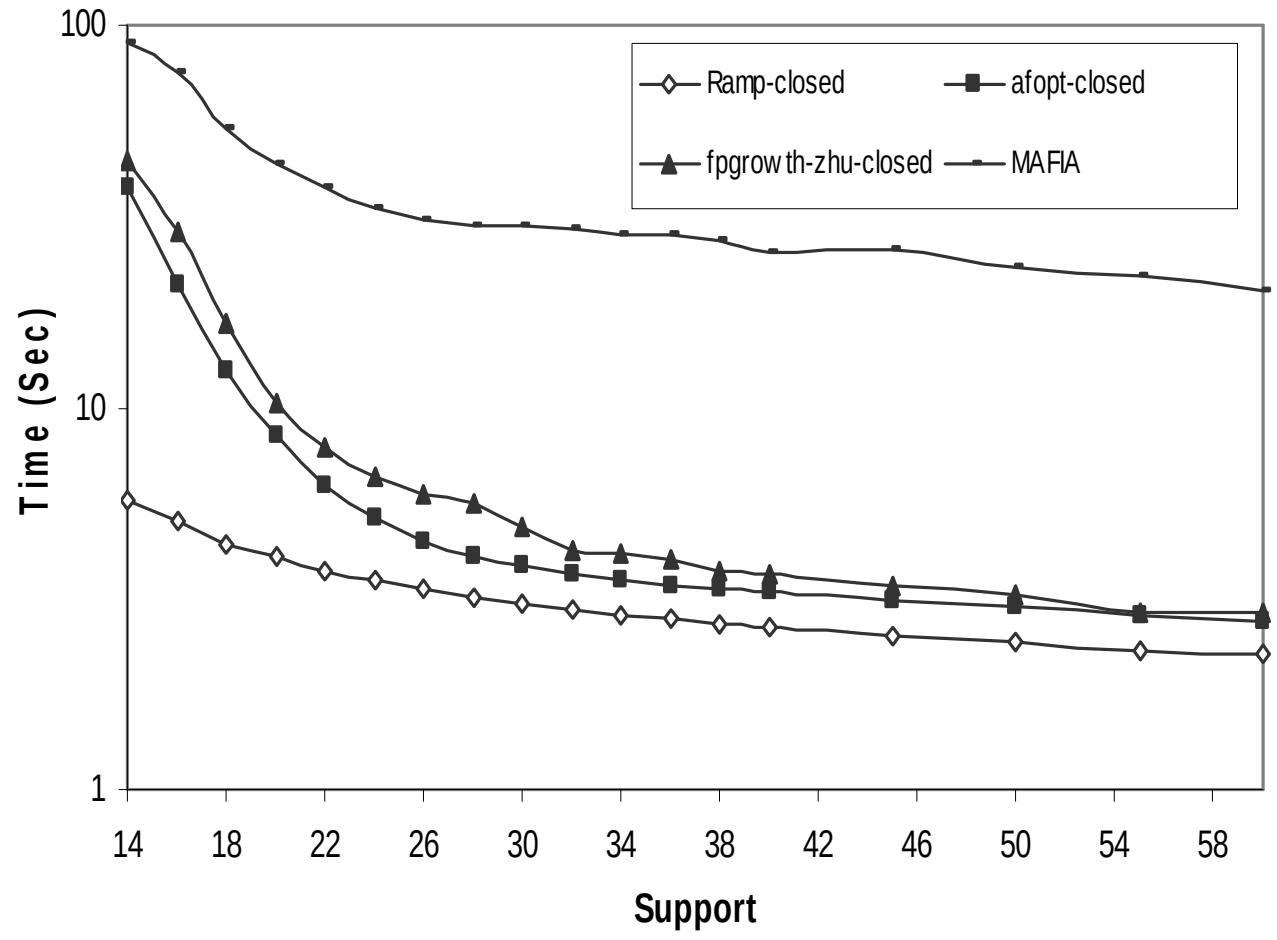

Figure 39: Performance results of Ramp-closed on T10I4D100K dataset.

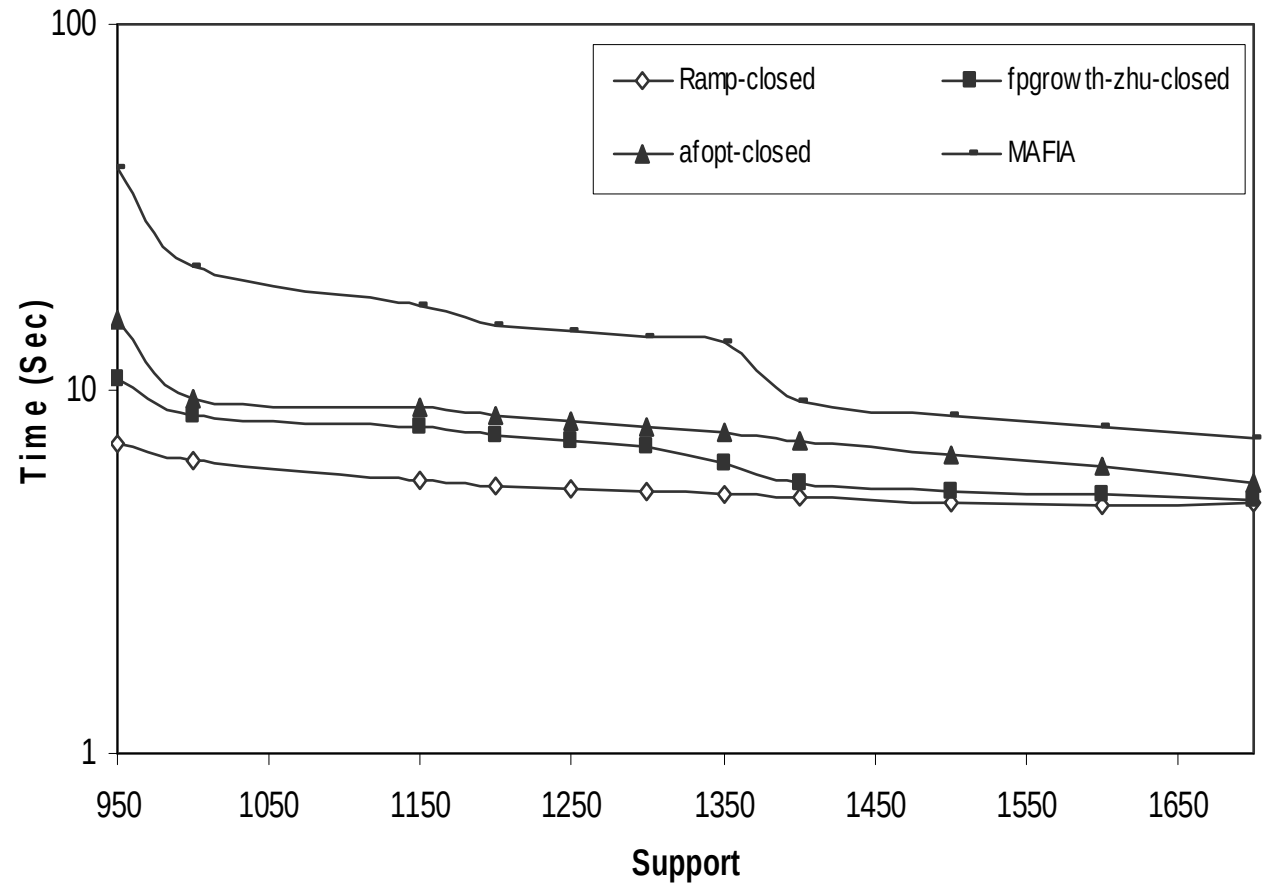

Figure 40: Performance results of Ramp-closed on T40I410D100K dataset.

## 8.3 Performance Analysis of Ramp-(all/max/closed) versus other Algorithms

Figure 19 to Figure 40 show the performance curves of all algorithms. The performance measure is the execution time of the algorithms datasets with different support thresholds. As we can see from Figures, the Ramp-all, Ramp-max and Ramp-closed outperform the other algorithms on almost all types of datasets, and give global best performance. The performance improvements of Ramp-all, Ramp-max and Ramp-closed over other algorithms are significant at low support thresholds.

## 8.4 Effect of FastLMFI on Ramp-max Implementation

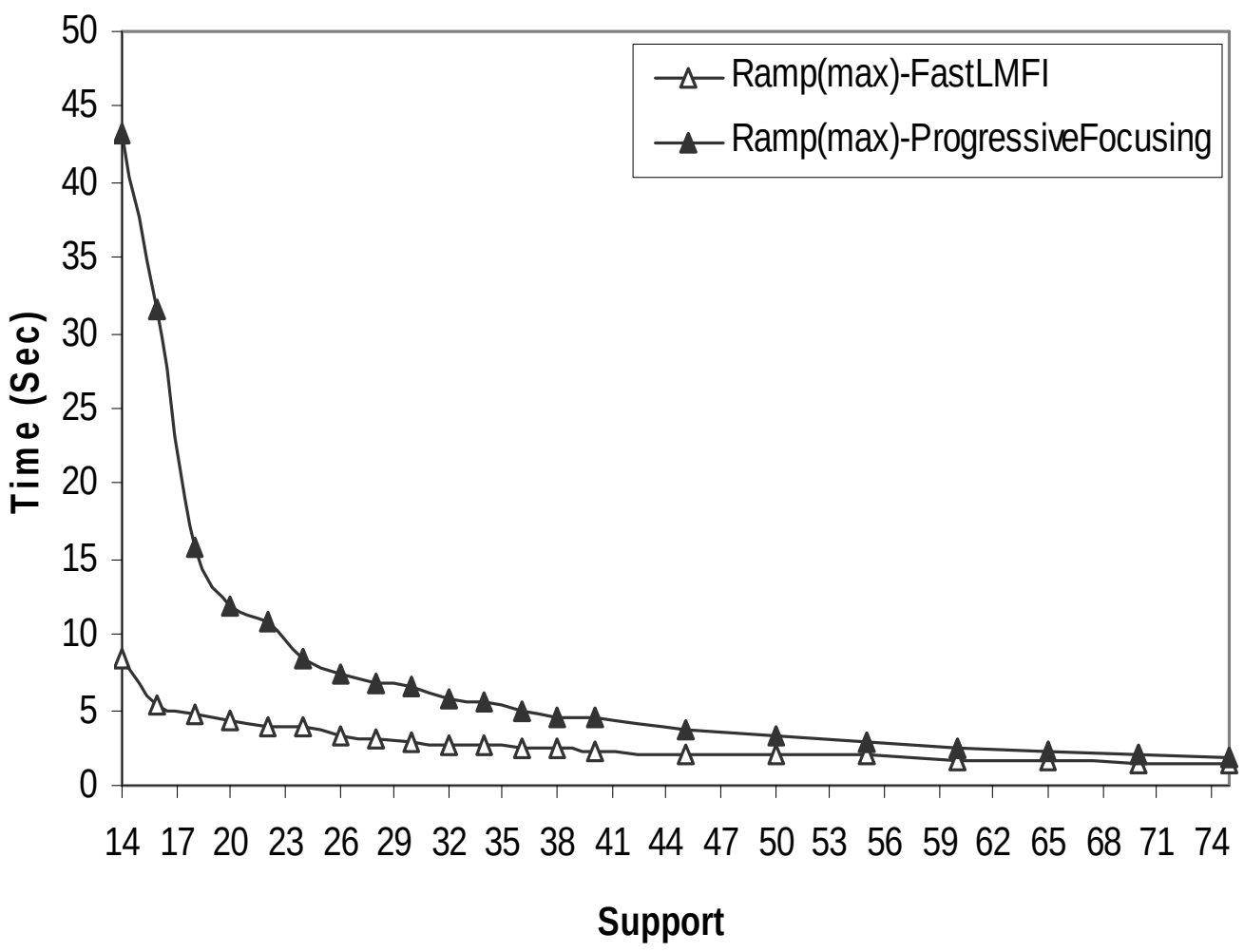

Figure 41: Effect of Ramp-max-FastLMFI on Retail dataset.

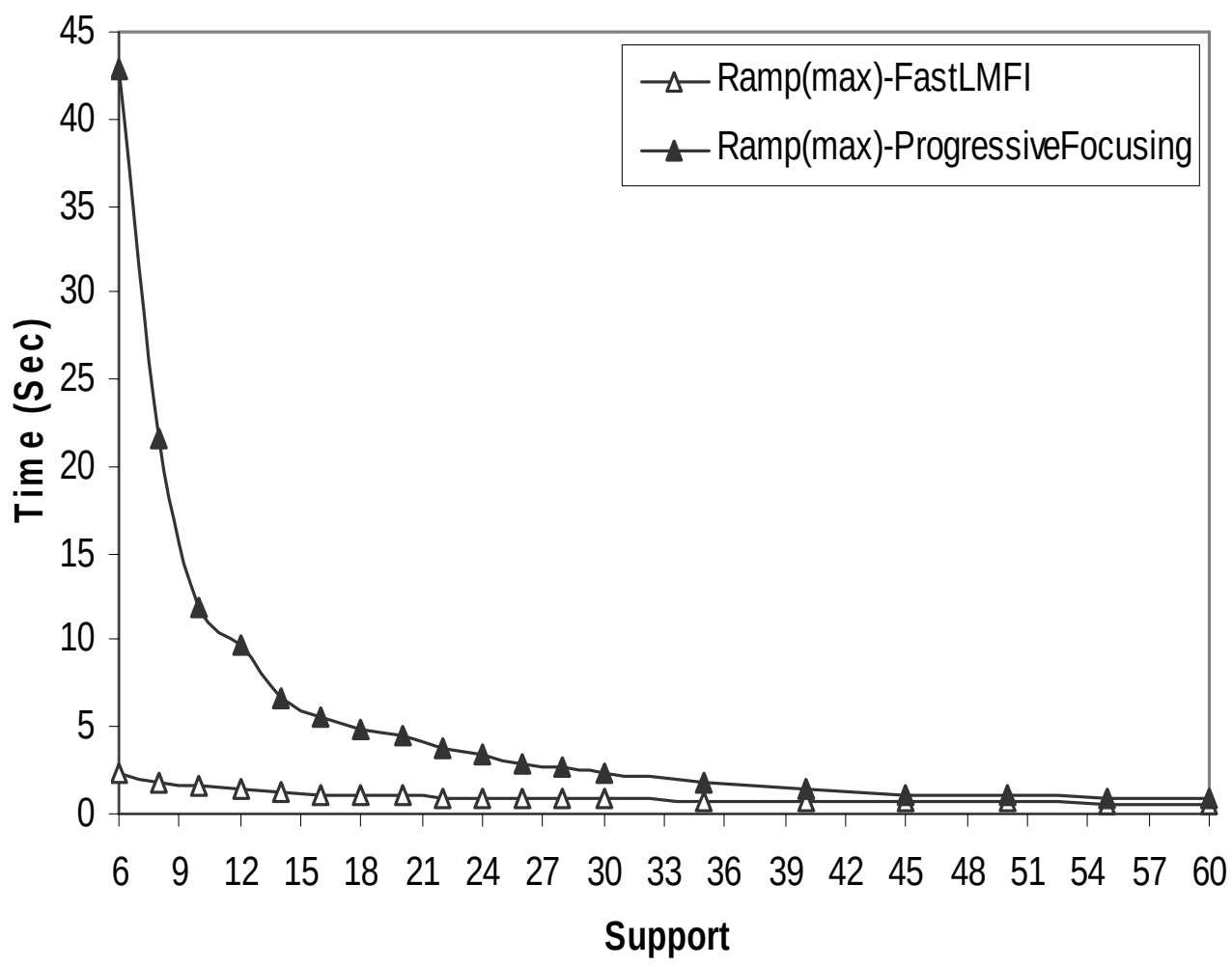

Figure 42: Effect of Ramp-max-FastLMFI on BMS-WebView2 dataset.

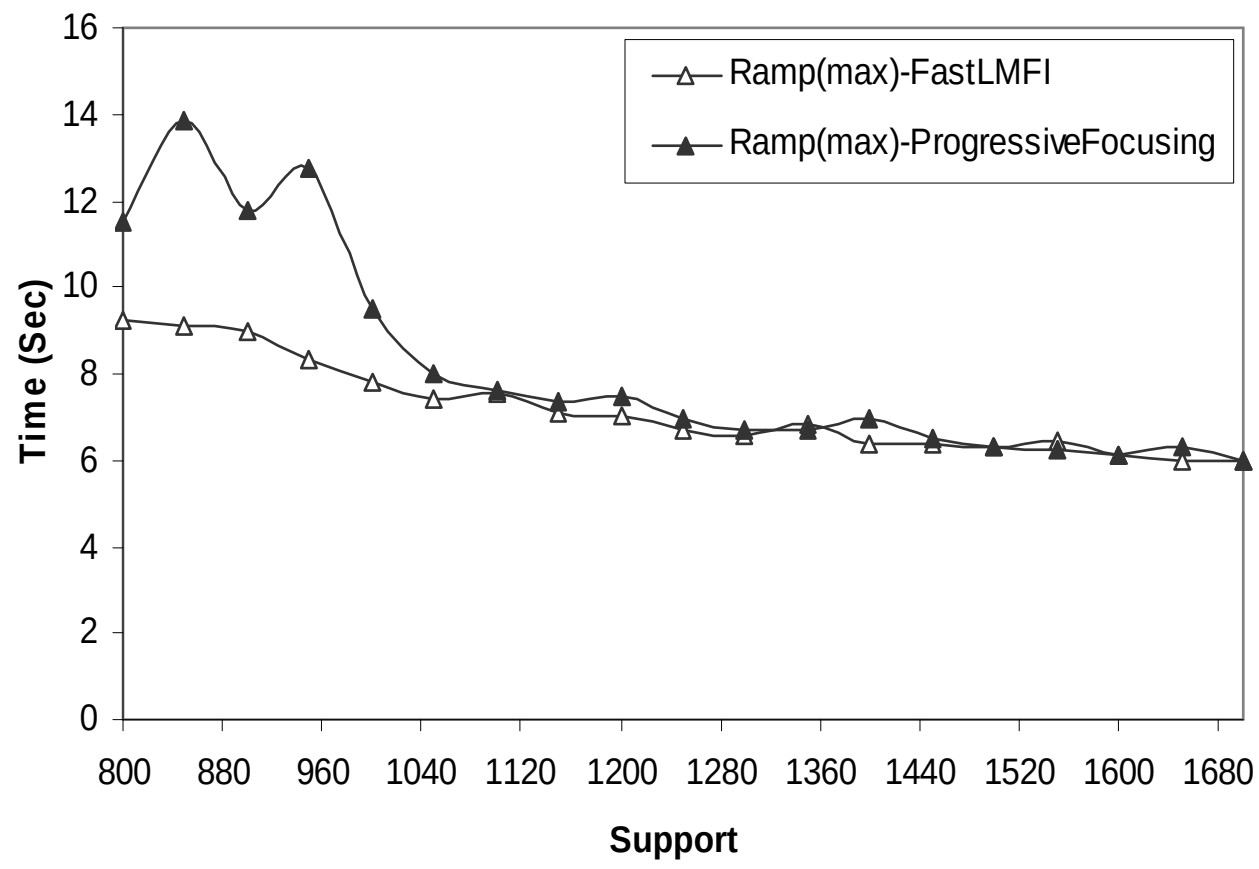

Figure 44. Effect of Ramp-max-FastLMFI on Mushroom dataset.

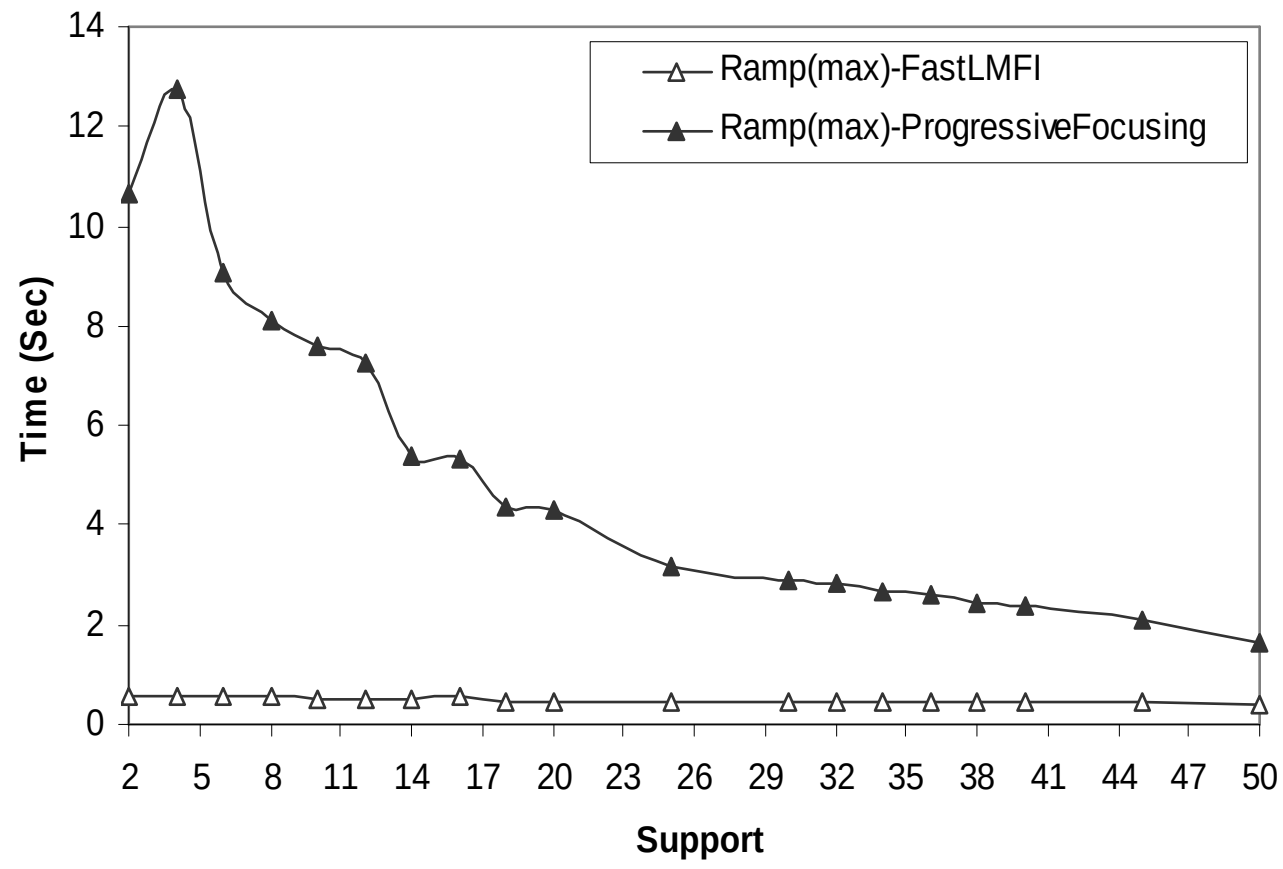

Figure 43. Effect of Ramp-max-FastLMFI on T40I410D100K dataset.

In this section we explained the effect of **FastLMFI** maximal frequent itemset superset checking on Ramp-max algorithm. The performance measure is the execution time of the Ramp-max-progressive focusing versus Ramp-max-FastLMFI on different support thresholds. For experimental analysis we used four datasets; Retail, BMS-WebView2, T40I4D100K and Chess. Figure 41 to Figure 44 are showing the performance curve of the two techniques. As we can seen from figures, checking itemset maximality by using **FastLMFI** is better than progressive focusing on almost all types of datasets.

## 9 CONCLUSIONS

Mining frequent itemset using bit-vector representation approach is very efficient for dense datasets, but highly inefficient for sparse datasets due to lack of any efficient bit-vector projection technique. In this paper we present a novel efficient bit-vector projection technique, which is better than the previous projected bitmap projection technique. The main advantages of our bit-vector projection technique are that, it does not require any rebuilding threshold or does not depend on any adaptive approach for projection, and can be easily applicable on all nodes of search space. We also present a new frequent itemset mining algorithm Ramp using our bit-vector projection technique. Different experiments on sparse as well as dense datasets show that Ramp is faster than the current best (all , closed and maximal) algorithms, which show the effectiveness of our bit-vector projection technique.